# Morphology-driven Optimization of Double Nanoholes for efficient and sensitive Plasmonic Optical Trapping

Pau Molet[1,2], Edona Karakaçi[1], Maria Sanz Paz[1,3], Mariano Barella[1,*], Michael Mayer[1,*]

[1]Adolphe Merkle Institute, University of Fribourg, Chemin des Verdiers 4, CH-1700 Fribourg, Switzerland.

[2]Present address: Department of Electrical Engineering, Stanford University, Stanford, CA 94305, USA.

[3]Present address: Sorbonne Université, CNRS, Institut des NanoSciences de Paris, INSP, F-75005 Paris, France.

*Corresponding authors:

mariano.barella@unifr.ch

michael.mayer@unifr.ch

## Abstract

Plasmonic Optical Tweezers based on Double Nanohole (DNH) structures are an emerging tool for trapping and interrogating single proteins under physiologically relevant conditions without the need for labeling or tethering. Nevertheless, their current performance is hindered by low signal-to-noise ratios for small proteins, fabrication variability, and photothermal instabilities, in part due to the required high laser power. To address these limitations, we present a comprehensive optimization of DNH parameters using parametric simulations and morphological characterization of experimentally fabricated DNHs. This approach ensures that the optimized geometries are experimentally realizable and compatible with established fabrication processes. We evaluate critical structural features, including gap width, gap length, cusp curvature, wedged tapers, adhesion layer and gold layer thicknesses, and the inclusion of pillars inside the nanoholes. By tailoring these variables, we aim to maximize the transmission variation upon trapping single nanoparticles or proteins and the local electric-field enhancement, thereby maximizing the optical force, while minimizing the required optical power. Two optimized DNH designs are proposed and simulated using a heuristic approach and a differential evolution algorithm. Both substantially outperform the reference structure and deliver 6.5- and 22-fold increases in sensitivity and 2.4- and 1.4-fold improvement in electric-field enhancement, respectively. These refinements provide a framework for developing efficient nanostructures to study single-protein dynamics with Plasmonic Optical Tweezers.

## Introduction

In recent years, Plasmonic Optical Tweezers (POTs) have emerged as powerful tools for single-particle studies, enabling trapping and interrogation of nanoscale objects[1–3]. In this technique, a dilute particle solution is deposited onto a substrate that presents plasmonic nanostructures. Under resonant illumination, strongly localized electromagnetic near fields are generated at the hotspot of the plasmonic nanostructure, producing optical forces capable of trapping nearby particles (Figure 1) and enabling the study of their dynamics[4]. Upon trapping, the particle locally modifies the dielectric function within the hotspot, shifting the plasmonic resonance and thereby changing the transmitted optical signal at a fixed wavelength. The same mechanism reports subsequent conformational changes of the trapped object. In contrast to conventional optical tweezers, the strategy of confining light to sub-wavelength volumes and enhancing the electromagnetic field enables efficient trapping and manipulation of small particles, such as proteins, DNA, and nanoparticles, with forces in the piconewton range[5–8]. This capability has made POT a compelling platform for studying molecular interactions, kinetics, and dynamics in fields ranging from biophysics to nanotechnology[9–11].

Over the years, various nanostructures have been used in POTs, including Double Nanoholes (DNHs)[12], bowties[13], inverted bowties[14], C-shaped apertures[15,16], coaxial rings[17,18], double nano rings, plasmonic and dielectric dimers[19,20] and pyramid arrays[21]. Each of these designs has contributed to the development and understanding of POT technologies, offering unique advantages in trapping efficiency, sensitivity, manipulation, and control of single-particle behavior. For instance, C-shaped apertures offer a unique methodology to transport nanometric particles via polarization control[15]. Coaxial rings have the best heat dissipation properties and are polarization insensitive[18]. Inverted bowties have demonstrated high electromagnetic confinement and low modal volume[22]. Recently, inverse design has produced topology-optimized apertures that resemble a DNH surrounded by a micrometer-scale coupler that funnels surface plasmon polaritons inward to boost the field intensity, and enhance the absorption and emission of luminescent trapped objects[23,24]. While this structure effectively increases the trapping strength, such an extended footprint may not be ideal for the transmission-based readout we discuss in this work: the coupler spreads both the incident and collected light over a large area, so the transmission modulation from a single trapped particle would represent only a small fraction of the total detected signal.

Among these designs, the DNH structure has garnered significant attention for its ability to create strong plasmonic fields with a well-defined trapping potential. Figure 1a illustrates the optical trapping of proteins using a DNH. Figure 1b shows a sketch (top view) of a DNH and Figure 1f shows a Scanning Electron Microscopy (SEM) image. The unique geometry of the DNH enables efficient confinement of light within the nanoscale gap between the holes and real-time monitoring of non-fluorescent trapped proteins or particles[12]. Furthermore, DNHs offer an efficient way to dissipate the heat generated by light absorption, as they are milled on a continuous layer of thin gold. Experimental studies on DNHs irradiated at 1064 nm have shown that the local temperature increase is around 10 °C at a moderate laser irradiance of 2 mW/µm$^2$ and can lead to a temperature comparable to that found in physiological environments[25]. Another advantage is the availability of numerous fabrication methods for DNHs. For instance, focused ion beam (FIB) milling offers precise control over the position, orientation, surface density, and morphology of each nanostructure. Examples of DNHs fabricated by this method will be presented throughout this article. Alternatively, colloidal lithography provides a faster and more cost-effective fabrication approach, although at the cost of losing control over the position and orientation of the structures, and limited control over the gap shape[26].

So far, these characteristics have made DNHs an excellent candidate for sensing the dynamics of single proteins[10,27]. Nevertheless, several limitations hinder their broader application to the study of conformational changes of small proteins. For example, once trapped, the transmission variation due to protein or enzyme conformational changes can be orders of magnitude smaller than the signal variation over the entire trapping event[28–30]. In general, the inherent Brownian motion of the protein within the trap produces fluctuations on the transmitted signal that are of the same order of magnitude as the ones associated with conformational changes. The result is a poor signal-to-noise ratio that makes the identification of conformational dynamics difficult. Increasing the sensitivity of the trap, i.e., the magnitude of the change in transmission when the protein is trapped at a given wavelength, could overcome that limitation.

Long-term stability of DNHs represents another limitation of the technique. It is well established that nanostructures fabricated from gold thin films are sensitive to storage conditions[31], as small gaps may widen, leading to degradation of the plasmonic properties, ultimately compromising the desired optical forces. Furthermore, when trapping small proteins, the high laser irradiance required may result in an excessive localized temperature increase incompatible with physiological conditions[32–34]. This thermal effect raises the risk of morphological alterations of the DNH or even structural failure, particularly during long-term experiments such as conformational dynamics studies of single proteins. Such transformations of the DNH affect the transmission signal and may be confused with conformational changes. Possible strategies to circumvent this problem include increasing the gold thickness to improve the heat dissipation[35], using sub-nanometer adhesion layers for increased thermal stability[36], and a gold film with large gold grains, as lower grain density means fewer diffusion channels[37].

In this article, we present a *morphological* parametric study of DNH performance using Finite-Difference Time-Domain (FDTD) simulations. First, we will provide the theoretical framework for the trapping mechanism in POTs, outlining the main equations governing trapping and the variation of the transmitted optical signal. Second, we analyze the optical properties of an actual DNH design used in previous single-particle trapping experiments from our group[10,27]. Then, we conduct a comprehensive series of simulations to identify the impact of each parameter in DNH-based POT. Finally, we propose an optimal set of parameters using criteria to reduce the laser power required for trapping while increasing the change in the transmitted optical signal upon trapping a protein-like particle.

We use the term *morphology* to denote the full set of geometric and layer-thickness parameters that define the as-fabricated DNH. These comprise the thin-layer thicknesses and the following structural features: gap width, gap length, cusp convexity, gap tapering, hole diameter, and the inclusion of pillars inside the nanoholes. To demonstrate that these geometries can be fabricated using standard techniques such as FIB, we provide SEM images of nanostructures produced with those parameters. Fabrication considerations, together with the morphological characterization of the fabricated structures, are mentioned throughout the text and detailed in the Supporting Information.

## Theoretical framework

The optical force exerted on a trapped object can be calculated as in conventional optical tweezers[38]. For particles smaller than the wavelength of the light, the Rayleigh approximation can be applied, and the total optical force is decomposed as the sum of three components accounting for the gradient force, the radiation pressure force, and the spin curl force[39]. In the following, we consider a plane-wave excitation, thus neglecting the last component. Moreover, for dielectric non-absorbing particles, the gradient force governs over the radiation pressure force, and the total force can be approximated by[40,41]:

$$F_{\text{trap}} \approx \frac{1}{4}\text{Re}(\alpha)\nabla|E|^2 \tag{1}$$

where α is the polarizability of the trapped object and $|E|^2$ is the squared electric field magnitude at the position of the particle, which is proportional to the local intensity. In the Rayleigh regime, particles and proteins can be modeled as a dipole, where, in the case of spherical objects, their polarizability is proportional to their volume, implying that small objects are subjected to weak optical forces[38,42]. However, in POT, the presence of a plasmonic nanostructure, such as a DNH, can exceptionally enhance the electric field in Eq. 1 if its wavelength is close to the nanostructure's resonance[43]. Figures 1c and 1g show the enhanced electric field in the region between the DNH cusps (inset of Figure 1a) . Under these conditions, the near field is tightly confined to sub-diffraction-limited volumes, generating an intense gradient force capable of trapping low-polarizability macromolecules without requiring laser intensities that can induce photodamage. This approach allowed the trapping of single proteins with laser powers in the mW range[3,10,27,28,44,45].

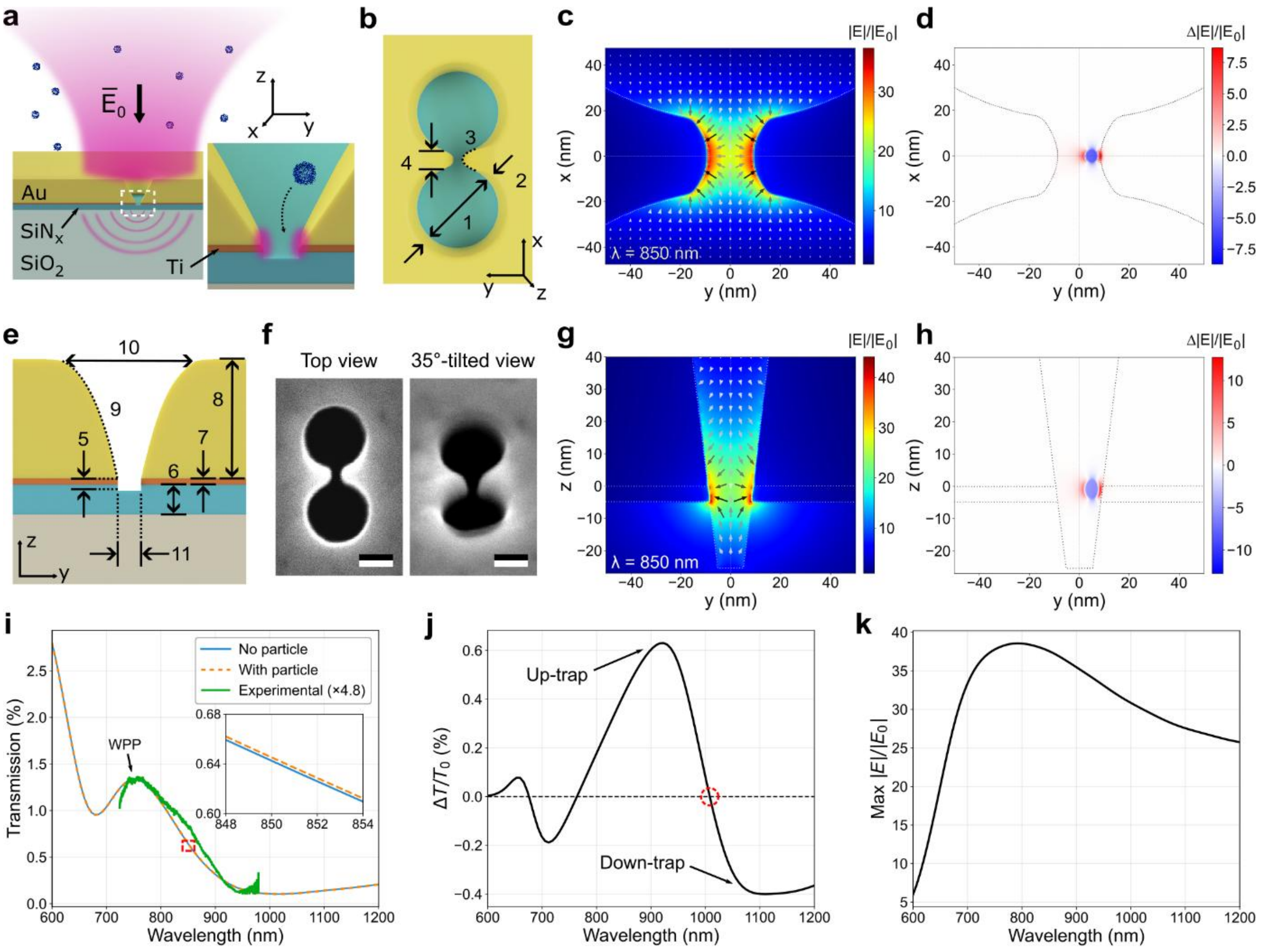

*Figure 1: Principle of Plasmonic Optical Tweezers. (a) A laser beam is tightly focused on a gold film with an engraved plasmonic Double Nanohole (DNH). The film blocks most of the incident light while the DNH acts as a diffraction-limited aperture that partially transmits the electric field. A protein solution is dispensed onto the gold substrate. Inset: A close-up sketch displaying the hotspot of the enhanced electric field. (b and e) Top view (xy) and cross-section view (yz) sketches indicating the relevant parameters in DNH morphology, also described in Table 1. Parameter 1: Bottom hole diameter. 2: Top hole diameter. 3: Cusp convexity. 4: Gap length. 5: Etch depth. 6: $SiN_x$ thickness. 7: Ti thickness. 8: Au thickness. 9: Tapered-gap wall profile. 10: Top gap width. 11: Bottom gap width. (c and g) In-plane top view (xy) and cross-section view (yz) of the electric field in the region between the cusps, normalized by the incident electric field. The 2D field plots show the electric-field enhancement at 850 nm. The optical force acting on a particle that sits inside the gap is represented by black and white arrows (for better contrast). The length of the arrow indicates the magnitude of the force as defined by Equation 1. (d and h) SIBA effect: Top view (xy) and cross-section view (yz) of the electric-field enhancement difference, that is, the electric field of a DNH with a particle at the hotspot minus the one of an empty DNH. Upon trapping, the particle plays an active role and locally increases the electric-field enhancement, thereby improving trapping efficiency. (f) SEM images of a representative DNH. (i) Experimentally measured and simulated Transmission spectra of the DNH shown in panel f. The measured spectrum was scaled by a factor of 4.8. See Table 1 for details on the geometry of the DNH from which this spectrum was acquired. The WPP is indicated with a black arrow. Inset: zoom-in of the region enclosed by the red rectangle. (j) Transmission variation upon trapping. The red dashed circle indicates the zero-crossing point between the up-trap and down-trap regimes. (k) Maximum electric-field enhancement at the hotspot as a function of the wavelength.*

In DNH-based POT, single-particle trapping is detected by monitoring changes in the light transmitted through the nanoaperture. When the nanoparticle/protein is trapped at the hotspot of the gap, it alters the dielectric function within the localized plasmon's mode volume, inducing a frequency shift in the plasmonic resonance. This shift introduces a variation in transmission at the excitation wavelength, enabling real-time tracking of the trapping process. Therefore, maximizing the frequency shift and the associated transmission variation upon trapping is essential for achieving high sensitivity. The expression for the frequency shift reads as follows[46]:

$$\delta\omega_0 = \omega_c \frac{\alpha}{2V_m\varepsilon_0} f \tag{2}$$

where $\omega_c$ is the cavity resonant frequency, $V_m$ is the modal volume of the plasmonic resonance, $\varepsilon_0$ the vacuum dielectric permittivity, and $f$ is the cavity intensity profile normalized to 1. The mode volume of a resonance can be calculated as[47]:

$$V_m = \frac{\int d^3x \frac{d}{d\lambda}(\epsilon(\lambda,x)\lambda)|E(x)|^2}{\max_x \left[\frac{d}{d\lambda}(\epsilon(\lambda,x)\lambda)|E(x)|^2\right]}, \tag{3}$$

where λ is the field wavelength and $\epsilon$ is the dielectric function of the media inside the gap without a particle. The numerator is the total electromagnetic energy stored in the mode, while the denominator is the peak energy density. Therefore, their ratio, $V_m$, has units of volume and measures the effective spatial extent over which the mode's electromagnetic energy is concentrated.

Early DNH-based plasmonic optical tweezers relied on the Fabry-Pérot gap mode (FP) – a standing-wave plasmon resonance across the hole diameter – as the operative trapping resonance[6,48]. While effective, this mode distributes the electromagnetic energy over a relatively large volume, limiting the achievable electric field gradients. Gordon's group identified a distinct resonance in tapered DNH gaps: the wedge plasmon polariton mode (WPP), which sits spectrally as a red-shifted shoulder to the FP resonance and exhibits a characteristic Fano line shape due to their interference[7,49] (Figure 1i). The wedge mode arises because gap surface plasmons propagating along the tapered gap walls, from the wide opening at the top towards the narrow bottom, are adiabatically slowed and squeezed as the gap narrows, concentrating electromagnetic energy into a sub-nanometric hotspot at the floor of the gap[50]. Compared to the FP mode, the WPP mode achieves significantly stronger field enhancement and a smaller mode volume at the same incident power. The WPP hotspots are depicted in the inset Figure 1a and present in the DNH structures simulated in the work presented here when excited at 850 nm, as shown in Figures 1c and 1g.

The WPP resonance presents as a better candidate for designing POT using DNHs, as the steeper intensity gradient generates a larger optical force on the target particle, enabling stable trapping at lower laser powers and reduced photothermal heating. Furthermore, the smaller modal volume amplifies the resonance frequency shift upon trapping (Eq. 2) and, consequently, the change in transmission, thereby improving detection sensitivity.

Based on these observations, we identify three key variables for optimizing the DNH design. First, the electric-field enhancement, $|E|/|E_0|$, which determines the optical force attainable at a given input power: higher enhancement yields a stronger local field and, consequently, a greater optical force without increasing the incident laser power. Second, the change in the transmitted signal resulting from a trapping event. Since the DNH is excited at a fixed wavelength during a trapping experiment, the resonance shift induced by the trapped object manifests as a change in the transmitted signal, making this quantity a direct measure of detection sensitivity. Third, the mode volume of the plasmonic resonance, which influences not only the magnitude of the transmission change via Eq. 2 but also the steepness of the field gradient – a smaller mode volume produces larger gradients and therefore stronger optical forces.

Herein, we present a series of simulations on tapered DNH structures with different morphologies. For each series, we discuss the optical properties considering the electromagnetic field enhancement at the WPP hotspot, the transmission variation upon trapping, and the mode volume of the plasmonic resonance.

## Simulation setup

The simulations are performed in the near infrared between 600 nm and 1200 nm, where the optical absorption of water and biological molecules is minimal compared to telecom and visible wavelengths. The morphological parameters we will explore herein are described in Table 1 and indicated in Figures 1b and 1e. The dimensions of the DNH structure were measured from SEM images of the as-fabricated nanostructure shown in Figure 1f and are listed in Table 1. The DNH fabrication procedure was previously described by our group[10] and is summarized in the Methods section.

*Table 1: DNH morphological parameters relevant for this work. These values were used to simulate the optical properties of the DNH presented in Figure 1.*

| Parameter | Value | Description |
|---|---|---|
| $\boldsymbol{t}_{\mathrm{Au}}$ | 100 nm | The thickness of the gold layer |
| $\boldsymbol{t}_{\mathrm{Ti}}$ | 5 nm | The thickness of the titanium adhesion layer |
| $\boldsymbol{t}_{\mathrm{SiN}}$ | 30 nm | The thickness of the silicon nitride layer |
| $\boldsymbol{g}_{\mathrm{w}}^{\mathrm{bot}}$ | 18 nm | The gap width at the bottom, that is, the distance between the two cusps at the gold/substrate interface |
| $\boldsymbol{g}_{\mathrm{w}}^{\mathrm{top}}$ | 100 nm | The gap width at the top, that is, the distance between the two cusps at the gold top surface. $g_{\mathrm{w}}^{\mathrm{top}} \geq g_{\mathrm{w}}^{\mathrm{bot}}$ due to the tapered profile |
| $\boldsymbol{g}_{\mathrm{l}}$ | 35 nm | The gap length, that is, the space separating the two nanohole edges |
| Cusp convexity | $a$ = 0.0130 | The projected cusp shape onto the $xy$ plane. Its sharpness or roundness is modeled as the coefficient $a$ of the function $y = ax^2$ |
| Wall taper | $b$ = 1 (linear) | The cross-section profile of the tapered gap at the cusps is modeled as the function $y = x^b$ where $b$ is a positive number |
| $\boldsymbol{\phi}_{\mathrm{bot}}$ | 190 nm | The bottom diameter of each nanohole at the gold/substrate interface |
| $\boldsymbol{\phi}_{\mathrm{top}}$ | 300 nm | The top diameter of each nanohole at the gold top surface. It's linked to the bottom diameter and the tapered profile ($\phi_{\mathrm{top}} \geq \phi_{\mathrm{bot}}$). It will not be directly explored |
| Etch depth | 20 nm | The substrate etch depth beyond the gold layer. First, etching the titanium layer, then the silicon nitride layer, and finally, the silica substrate |
| $\boldsymbol{\phi}_{\mathrm{pil}}$ | None | The presence of a pillar inside each nanohole |

We performed simulations to compute the transmission spectrum $T$ and the electromagnetic field $|E|$ around the DNH hotspot, in water. Additionally, we mimicked the presence of a trapped protein by placing a spherical particle with a diameter of 6 nm in the DNH gap (see Methods for further details). We then computed the transmission spectrum of the DNH-particle system, hereinafter referred to as $T_{\mathrm{trap}}$. Next, we calculated the transmission variation upon trapping, defined as $\Delta T/T_0 = 100 \cdot \frac{T_{\mathrm{trap}} - T_0}{T_0}$, where $T_0$ denotes the transmission spectrum in the absence of the particle. We will also refer to this quantity as *sensitivity*, as it allows us to quantify how responsive a given nanostructure is to a trapping event.

Figure 1i shows the simulated and measured transmission spectra of the DNH shown in Figure 1f. The nanostructure dimensions and morphology are such that the WPP resonance falls within the wavelength range of interest. Notably, the measured transmission spectrum of the DNH captures this feature. The agreement between the simulation and the experiment extends across the entire measured spectral range, indicating that the modeled morphology closely represents the as-fabricated DNH. The measured spectrum was scaled vertically by a factor of 4.8 to facilitate comparison with the simulated data (see Methods for further details).

Upon trapping, the spectrum appears unchanged at first glance. However, a slight red shift can be discerned in the inset of Figure 1i. The spectral change becomes evident in Figure 1j, which displays the transmission variation upon trapping, $\Delta T/T_0$. This curve exhibits two extrema of high sensitivity: a maximum at 910 nm and a minimum at 1100 nm. These spectral features define two distinct trapping regimes, separated by the zero-crossing point (indicated by a red circle in Figure 1j). Depending on the trapping laser wavelength relative to this point, two types of trapping signatures are observed, named up-trap and down-trap events. Specifically, when the laser wavelength is blue-shifted (red-shifted) with respect to the zero-crossing point, trapping results in an increase (decrease) of the transmitted signal.

Figure 1c shows the in-plane ($xy$) electric-field enhancement distribution $|E|/|E_0|$ at the interface between the Ti adhesion layer and the silicon nitride ($z$ = -5 nm), computed at a wavelength of 850 nm, where $|E_0|$ denotes the incident electric field amplitude. Figure 1g shows a cross-sectional ($yz$) view of the enhancement at the narrowest point of the DNH gap ($x$ = 0). The hotspots are located between the cusps and span the narrowest region of the gap, extending vertically through the adhesion layer and a few nanometers over the gold wall of the gap, reaching a maximum field enhancement of 43. The gradient force was computed using Eq. 1 and is represented by arrows in Figures 1c and 1g. Given that the polarizability of the protein-like particle is real and positive, the resulting optical forces on a particle entering the gap are directed toward the cusps, indicating that the particle will be driven toward the hotspots.

Figure 1k shows the maximum field enhancement in the hotspot region as a function of the wavelength. At 850 nm, the maximum enhancement is ~37, slightly lower than the value quoted above. This difference arises from the coarser mesh used to reduce the computational cost of the broadband simulations, which averages the field over a larger volume, thereby reducing the peak value (see Methods for details). The field enhancement remains above 25 across most of the NIR range, covering both the up-trap and down-trap regimes, with a peak value of ~39 at 791 nm. This observation indicates that, in principle, trapping is feasible at virtually any wavelength within this range. However, operating far from the sensitivity extrema would compromise the ability to detect a trapping event.

The presence of the trapped object modifies the electric field distribution in the trap. Figures 1d and 1h display the difference in amplitude of the electric field normalized by the amplitude of the incident electric field, i.e., $\Delta|E|/|E_0|$, where $\Delta|E| = |E_{\text{trap}}| - |E_{\text{no trap}}|$. Around the particle, the local electric field is even more enhanced, about 25%, than that of the empty gap. Moreover, because the particle was placed closer to one of the cusps (see Methods), the field is slightly higher on that side. Changing the particle position leads to lower enhancements, but they remain higher than in the empty case (see Supporting Information Section S1). These observations evidence the active role of the particle in the plasmonic optical trapping mechanism, the so-called Self-Induced Back-Action (SIBA) effect[48]. Quidant's group proposed that down-traps or resonant-traps lead to unstable trapping, while up-traps offer improved power-input efficiency as the SIBA mechanism helps to reduce the necessary input power for trapping[46].

Before proceeding to the results, we briefly discuss the explored parameter range. While many studies have simulated DNHs with straight gaps[6,25,51], such geometries (zero convexity and 90° sidewall slopes) do not accurately represent actual fabricated nanostructures. As shown in Supporting Information Section S2, DNH structures with both short and long gaps exhibit a curved cusp profile in the $xy$ projection. Similarly, FIB milling a structure with a straight wall shape is not feasible. Careful inspection of fabricated structures reveals rounded edges at the top surface of the gold film (see representative SEM images in Supporting Information Figure S4). This is a well-known effect of high-aspect-ratio nanostructures (e.g., 18 nm-wide gaps in 100 nm-deep trenches) fabricated using Gaussian focused ion beams[52]. Therefore, to provide insights for realistic, optimal nanostructure design, we will consider DNH structures with cusps whose convexity is greater than zero and have a curved wall shape.

## The plasmonic and support layers

### Gold layer

The first parameter to explore is the thickness of the plasmonic film. Figure 2a shows the transmission spectra for thicknesses varying from 60 nm to 200 nm with 10 nm increments. As expected, the thinner the gold film, the higher the transmission. However, films thicker than ~150 nm exhibit strong attenuation in the visible range and slightly increased transmission above 800 nm. Figure 2b shows that $\Delta T/T_0$ blueshifts as the thickness increases, and so does the up-trap/down-trap transition (Figure 2o). Coupled with that, up-trap and down-trap peak values of $\Delta T/T_0$ increase with thickness (Figure 2p), representing a straightforward way to tune the sensitivity of the DNH for the desired or available laser wavelength. Apart from this improved sensitivity, Figure 2c shows that thicker layers also yield greater electromagnetic field enhancements, resulting in stronger forces, and offer better heat dissipation due to an increased thermal conductivity[35,53].

Increasing the gold thickness, however, comes at the expense of an enlarged mode volume. As shown in Figure 2d, while more energy is stored in the thicker structure, the spatial confinement of the field is reduced as the mode volume increases. This effect results in moderate field gradients at the DNH cusps, which, although sufficient to generate optical forces capable of trapping nanometer-scale objects, are not optimal. Nevertheless, the trapping energy, defined as the integral of the electromagnetic energy density over the volume occupied by the particle, indicates an optimal gold thickness of approximately 170 nm.

An additional consideration not captured by the simulations is the grain size of the gold film. Small grains not only promote structural instabilities but also complicate the fabrication process. Gold films deposited at high temperatures by electron-beam evaporation typically exhibit large grains, as shown in Supporting Information Section S3. Sufficiently large grains can entirely contain the cusp region of the DNH during FIB milling, thereby providing greater structural integrity to the fabricated nanostructure.

### Titanium layer

The material and thickness of the adhesion layer are crucial for preserving the characteristics of the deposited gold in thin-film plasmonic devices. Numerous studies have shown that the lossy behavior of commonly used metals, such as Ti or Cr, reduces the quality factor and increases the substrate's local temperature[34,54,55]. Here, we focused on titanium adhesion layers with thicknesses ranging from 0 (none) to 9 nm in 1 nm increments. As shown in Figure 2e, increasing the thickness of the Ti layer decreases (slightly increases) the transmission below (above) ~775 nm. More importantly, when the Ti thickness is increased, the $\Delta T/T_0$ maximum value is reduced, compromising the sensitivity of the design (Figure 2f). Also, the transition between up-trap and down-trap behavior shifts towards the red as the Ti thickness increases, as displayed in Figure 2o. It also comes at the expense of a decrease in the electric-field enhancement (Figure 2g) and an increase in the plasmonic modal volume (Figures 2h), which lowers the Q-factor of the cavity. These observations suggest that the thickness of the Ti adhesion layer should be minimized when looking for an optimized DNH design.

Previous studies on the role of the adhesion layer report that the optimal thickness for optical and thermo-plasmonic applications is around 0.5 nm, while maintaining adhesion properties. At this sub-nanometer scale, the titanium forms a discontinuous film (island growth) rather than a continuous one, while preserving the crystallinity of the gold growth and drastically reducing plasmonic damping[56]. Therefore, reducing the thickness of the metallic adhesion layer or using nonabsorptive alternatives is advantageous. The community suggested the use of self-assembled monolayers like MPTMS[54], APTMS or APS[57], although the thiol bond between the organic molecule and the gold may be thermally unstable, limiting the temperatures of the gold deposition and the experiments[54] (see Supporting Information Section S3 for their use in film deposition).

**Silicon nitride layer**

Plasmonic optical tweezers using DNH have been used together with $SiN_x$ nanopore technologies for single-protein sensing and characterization. In this approach, called SIBA Actuated Nanopore Electrophoresis (SANE), the DNH structure is fabricated on top of the $SiN_x$ solid-state nanopore and used to trap the molecules of interest at the nanopore entrance, thereby counterbalancing the electrophoretic force[58,59]. Then, standard solid-state nanopore current-blockade measurements are performed while trapping the particle. This approach provides orthogonal readouts: the change in optical transmission confirms trapping and conformational dynamics, while the ionic current blockade yields precise sizing and electrokinetic information. In the context of this technique, it is relevant to discuss the effect of including a $SiN_x$ layer below the plasmonic layer and the substrate.

The simulations presented in Figures 2i and 2j, show that adding a $SiN_x$ layer causes a subtle redshift in the plasmonic resonance, which also appears in $\Delta T/T_0$ together with a reduction in the peak value. Moreover, thick $SiN_x$ layers significantly lower the maximum electric-field enhancement at the hotspots, reducing the optical force (Figure 2k). Additionally, the inclusion of a $SiN_x$ layer beneath the adhesion layer slightly increases the modal volume of the plasmonic hotspot, regardless of its thickness, as shown in Figure 2l. These observations suggest that adding an extra layer to the fabrication process is not beneficial if the DNH is not used for SANE experiments, as most relevant optical properties are adversely affected. Nevertheless, the advantage of incorporating a $SiN_x$ layer is that, when deposited as a non-stoichiometric, low-stress film, it can serve as a barrier against oxidation of the underlying Ti adhesion layer, thereby improving the overall robustness of the fabricated nanostructure. On fused silica substrates, oxidation of the Ti interlayer may compromise adhesion, leading to microcracks and eventual peeling of the Au film, as gold adheres poorly to titanium oxide[56].

As shown in Supporting Information Section S3, the presence or absence of the $SiN_x$ layer has a negligible effect on the gold grain size in the $SiO_2$/Ti/Au (190 °C) stack. This is advantageous, as the $SiN_x$ layer can be incorporated for its structural benefits without affecting the film morphology, as discussed above.

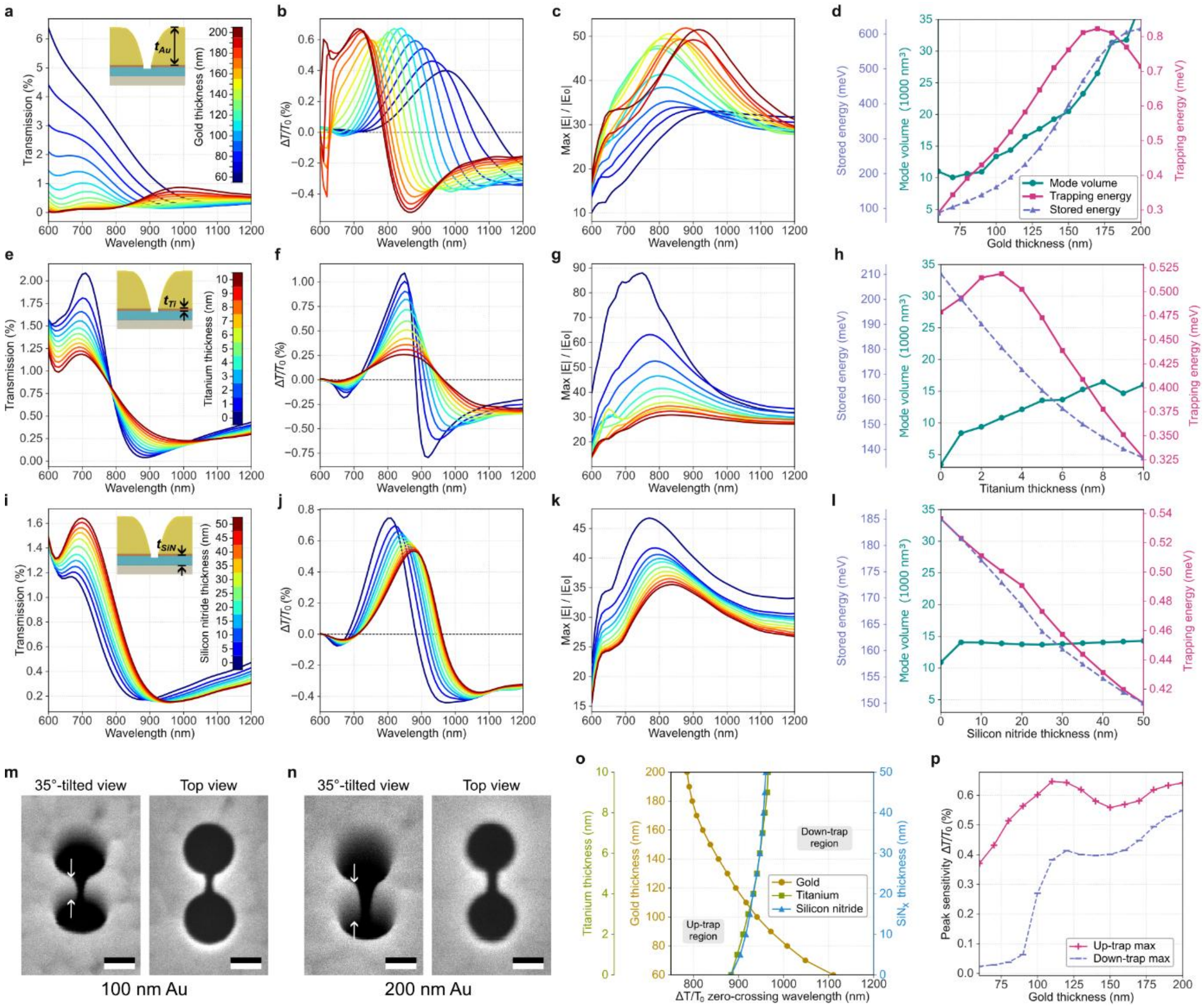

*Figure 2: The role of the thickness of the plasmonic and support layers on DNH optical properties. (a, e, and i) Transmission spectra for different Au, Ti, and $SiN_x$ thicknesses, respectively. (b, f, and j) Transmission variation upon trapping $\Delta T/T_0$ for different Au, Ti, and $SiN_x$ thicknesses, respectively. (c, g, and k) Maximum electric-field enhancement for different Au, Ti, and $SiN_x$ thicknesses, respectively. (d, h, and l) Stored energy, mode volume, and trapping energy for different Au, Ti, and $SiN_x$ thicknesses, respectively. (m and n) Top and tilted view of DNHs fabricated with 100 nm and 200 nm-thick gold, respectively. (o) $\Delta T/T_0$ zero-crossing wavelength as a function of the layers' thicknesses. (p) Peak sensitivity as a function of the gold thickness for up-trap and down-trap configurations. The peak sensitivity wavelength varies with thickness.*

# Gap morphology

## Gap width

The relevance of the gap width has been a subject of study in plasmonic sensing, as smaller plasmonic gaps produce stronger electric-field confinement. In particular, in DNH-based optical trapping, it was also shown that $\Delta T/T_0$ and the field enhancement increase dramatically with decreasing gap width[7,14,60,61]. The simulations performed here point in the same direction but further reveal that the width at the bottom of the gap significantly affects all studied optical properties of the DNH.

Figure 3a shows that increasing the gap width blueshifts and attenuates the WPP resonance while simultaneously increasing the transmission in the NIR range. Coupled with this effect, Figure 3b shows that the sensitivity peak also blueshifts (see also Figure 3p) and diminishes in magnitude, thereby reducing the overall sensitivity of the DNH trap. Nevertheless, arbitrarily small gaps are not desirable, as the size of the target protein or nanoparticle constrains the minimum accessible gap width.

As expected, larger gaps exhibit a reduced field enhancement (Figure 3c) as the increased separation between the opposing cusps diminishes the spatial overlap of the plasmonic modes. This observation is further supported by Figure 3d, where the mode volume increases with the separation of the cusps and the hotspots become progressively decoupled, diminishing their electromagnetic interaction.

Figure 3m shows that it is feasible to fabricate DNHs with gap widths as small as 13 nm using a $Ga^+$ FIB, while gaps as narrow as 4 nm have been achieved through proximal milling and helium FIB, albeit with potentially reduced reproducibility[14,22,62].

**Cusp convexity**

The cusp convexity directly impacts the electric-field enhancement. As the cusps get sharper, their "point effect" clearly becomes stronger and the DNH structure can tightly confine electromagnetic energy within very small volumes. To analyze the impact of this phenomenon, we simulated DNHs with $g_l$ = 40 nm and $g_w$ = 18 nm, while varying the convexity of the cusps. The shape of the cusps was modeled as a parabolic function $f(x) = ax^2 + c$, where $x$ is the spatial coordinate (see Figure 1b), $a$ defines the convexity via the second derivative $f'' = 2a$ and $c$ is $g_w/2$, where the perimeter of the cusp intersects the $x$ axis.

Figure 3e shows how the transmission spectrum changes as a function of the cusps' convexity. From almost flat to highly convex, the spectra exhibit a small blue shift, and so does $\Delta T/T_0$ (Figure 3f). As for the electric-field enhancement in Figure 3g, the "point effect" is evident: high convexity leads to greater confinement and larger enhancement. In DNHs with geometries characterized by low (high) convexity, the cusp hotspots extend over a large (small) portion of the gap space. This effect increases (reduces) the mode volume as the electromagnetic energy spreads out rather than being confined (Figure 3h).

From a fabrication standpoint, the convexity can be tuned by varying the milling time for each nanohole. Long times intensify the proximity effect, leading to low convexity. Also, adding subsequent milling steps along the edges of the gap reduces the convexity of the cusps.

**Gap length**

Figures 3i to 3l illustrate the role of the gap length. As the gap length increases, the transmission resonance undergoes a marginal blueshift, accompanied by a reduction in field enhancement. Several unfavorable trends are evident: sensitivity decreases, and field enhancement is not only reduced in magnitude but also less spatially confined as the mode volume increases.

It is worth noting that the gap length directly affects the convexity of the cusps. Specifically, longer gaps produce rounder cusps, as can be observed in the SEM images shown in Figure 3o. This connection was observed across several DNHs fabricated with varying gap lengths (see Supporting Information Section S2).

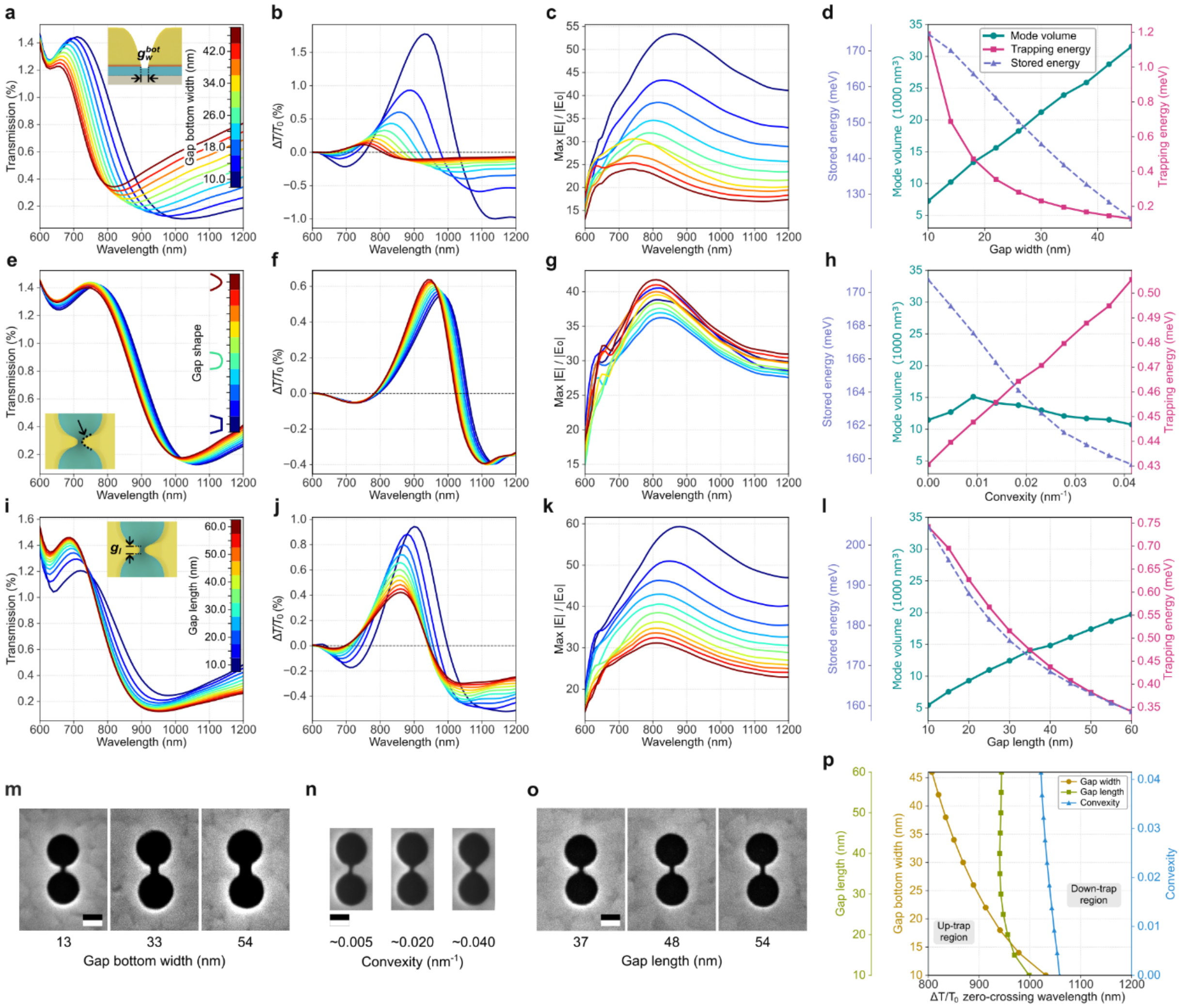

*Figure 3: The impact of DNH gap morphology on the optical properties. (a, e, and i) Transmission spectra for different gap bottom widths, gap shapes, and gap lengths, respectively. The shape is modified via the convexity of the cusps. (b, f, and j) The corresponding transmission variation upon trapping $\Delta T/T_0$. (c, g, and k) The corresponding maximum electric-field enhancement. (d, h, and l) The corresponding stored energy, mode volume, and trapping energy. (m, n, and o) SEM images (top view) of DNHs fabricated with different gap bottom widths, gap shapes, and gap lengths, respectively. (p) $\Delta T/T_0$ zero-crossing wavelength for the explored parameters.*

## Tapered gap: wall shape and top gap width

Tapered-gap structures feature inclined wall profiles to produce narrow 3D features, such as a sharp tip or a wedge. In these structures, the top part of the gap at the water/gold interface is wider than the bottom part of the gap at the titanium/silicon nitride interface. In this way, the wedge surface plasmon propagates along the $z$ direction from the top interface to the bottom one, focusing as the gap narrows[50,63]. Figure 4m shows several examples of DNH structures fabricated with different tapered profiles (see Methods for details on the fabrication process). They exhibit a curved wall profile that was modeled by a power function $f(x) = ax^b$ where $x$ is the spatial coordinate, $a$ is a factor that helps match the gap dimensions at the top water/Au interface with the top diameter of the holes, and $b$ is the taper power that shapes the wall profile defined by $b = 1/N$ where $N$ is a positive number.

Figures 4a to 4d show the results of simulations with varying taper power with $N$ ranging from 0.8 to 5. Transmission spectra are not significantly affected by the tapering. A slight overall increase of 0.2% to 0.4% is observed for highly tapered gaps ($N$~1). Significant changes in $\Delta T/T_0$ and field enhancement are not observed either, except for $N \leq 1$. The electric-field confinement exhibits a subtle trend. High tapering (small $N$) yields a small mode volume that monotonically increases as tapering reduces (large $N$). High $N$ produces structures resembling those with straight-walled profiles, thereby negatively affecting the focusing power of the propagating WPP within the gap. This effect reduces the enhancement, spreads the mode energy along the gap, and increases the mode volume, as shown in Figure 4d. In general, a strong taper allows for tight confinement of the electric field and energy without affecting the transmission or compromising the sensitivity.

Alternatively, one could modify only the gap width at the top interface to create a tapered gap. Figure 4n shows DNH structures with gap widths at the top of ~108, ~112, and ~121 nm, but with similar power of the wall taper. Figures 4e to 4h show the effect of varying the gap width at the top while using a $x^{1/N}$ tapered wall profile with $N$=1. The gap width at the top interface ranges from 18 to 190 nm with 20 nm steps (see Supporting Information Section S4 for matching hole top diameters). The gap width at the bottom interface was fixed to 18 nm. In contrast to the previous strategy, enlarging the width of the gap at the top significantly affects the optical properties. In the NIR, above ~800 nm, the transmission increases while the sensitivity and the enhancement redshift (see Figure 4o). At 850 nm, a standard wavelength used for trapping, gap widths at the top of 70-90 nm yield the highest sensitivity for this DNH morphology (Figure 4p). The mode volume and the energy reach their minimum and maximum, respectively, for gap top widths between 70 and 150 nm, yielding values similar to those obtained with a strong taper ($N$~1).

**Substrate etch**

We also studied the effect of the etch depth, which is controlled by the FIB milling time of the DNH gap area. Milling for extended periods etches the $SiN_x$ layer under the Ti, eventually exposing the fused silica beneath. On the contrary, short milling times can result in poor or even failed Ti etching.

We simulated DNH structures with different etch depths ranging from 2 nm to 60 nm. These limits are equivalent to having a 3 nm-thick layer of Ti within the gap (2 nm etch depth), and to fully etched Ti and $SiN_x$ layers with a 30 nm-deep trench in the fused silica substrate (60 nm etch depth). Figures 4i and 4j show the transmission and $\Delta T/T_0$ spectra of DNH structures with varying etch depths, respectively. No meaningful differences are observed between the simulated spectra for the explored etch depths. Figure 4o shows that the $\Delta T/T_0$ zero-crossing point essentially does not shift in such configurations. Instead, the electric-field enhancement increases with the removal of the Ti and $SiN_x$ layers. Additionally, a deeper trench at the gap provides more space for the trapped object to reach the hotspot than a poorly milled trench. Given that the mode volume is nearly constant at ~13000 $nm^3$, the particle can occupy regions with high electric fields, thereby enabling the trapped object to access higher energies, as shown in Figure 4l.

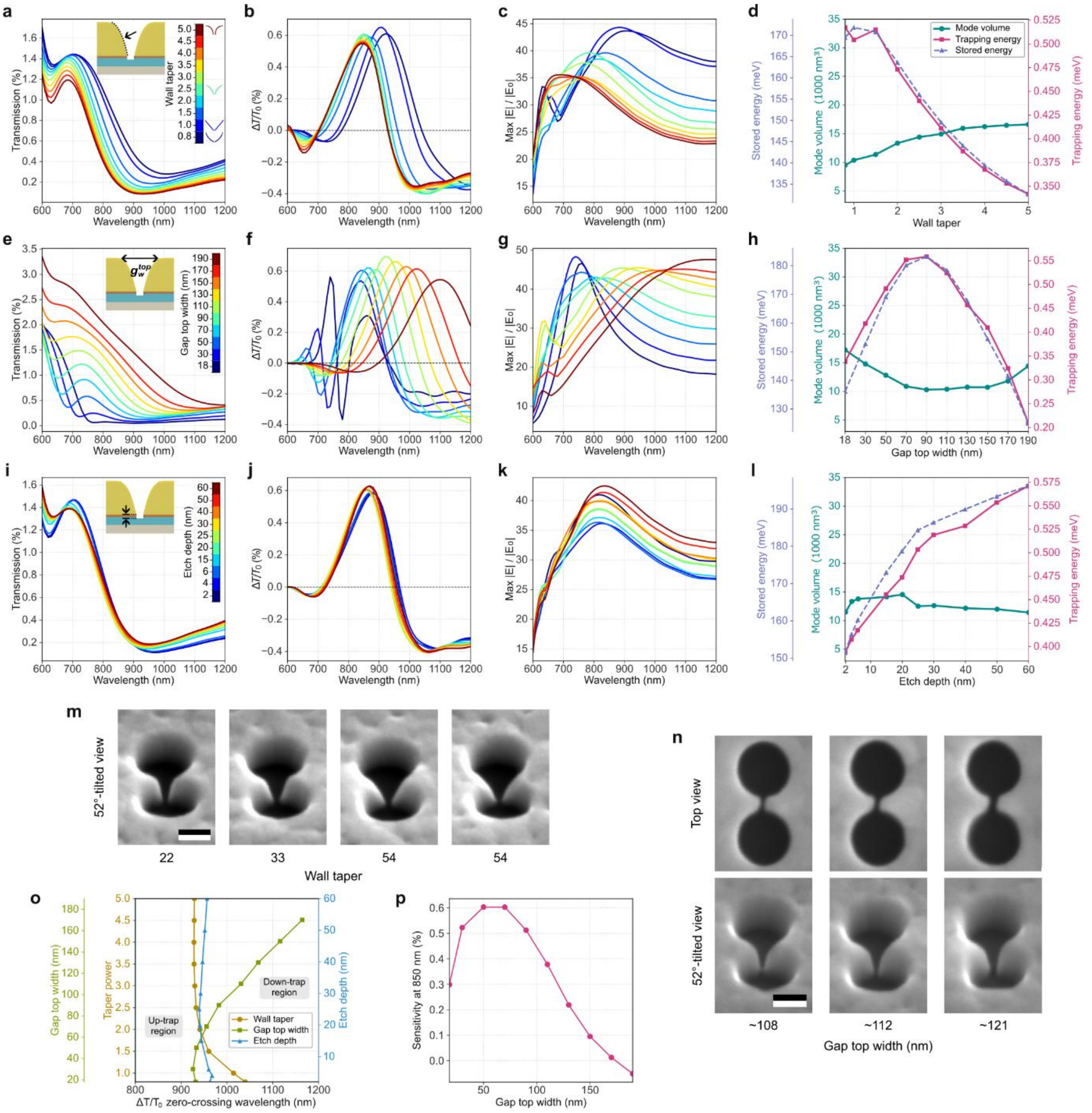

*Figure 4: The impact of DNH gap tapering and etching depths on the optical properties. (a, e, and i) Transmission spectra for different powers of the taper of the gap wall, gap top widths, and etch depths, respectively. (b, f, and j) Corresponding transmission variation upon trapping $\Delta T/T_0$. (c, g, and k) Corresponding maximum electric-field enhancement. (d, h, and l) The corresponding stored energy, mode volume, and trapping energy. (m) SEM images (52°-tilted view) of DNH structures fabricated with different tapered profiles of the gap wall. (n) SEM images (top and 52°-tilted views) of DNH structures fabricated with different gap widths at the top. (o) $\Delta T/T_0$ zero-crossing curves for the explored parameters. (p) Sensitivity at a trapping wavelength of 850 nm as a function of the gap width at the top.*

# Hole morphology

## Hole diameter

We simulated the optical properties of DNH structures by varying the diameter of the nanoholes at the bottom Au/Ti interface $\boldsymbol{\phi}_{\mathrm{bot}}$. The top diameter, $\boldsymbol{\phi}_{\mathrm{top}}$, was set to be 30 nm larger than the bottom diameters to account for the proximal etching effect of FIB fabrication. Figure 5i shows examples of DNH structures fabricated with 127, 150, 174, and 194 nm diameter holes.

Figure 5a shows the transmission spectra of different DNH structures with hole diameters ranging from 140 nm to 240 nm with 10 nm steps. Increasing the diameter of the holes redshifts the plasmonic resonance and increases the transmission magnitude, in a manner analogous to Bethe's law for far-field transmission through subwavelength apertures[64]. This trend agrees with early observations made by Prikulis *et al.* on single nanoholes milled into 20 nm-thick Au films[65], and with FDTD simulations by Jiang *et al.,* who showed that the transmission spectrum of DNH structures with relatively small diameters (60 to 130 nm) redshifts as the diameter increases[6]. Figure 5b shows the transmission variation upon trapping $\Delta T/T_0$ that exhibits a similar trend, which is also recognizable in the nearly linear relationship between the wavelength of maximum sensitivity and the hole diameter (Figure 5k). The electric-field enhancement also exhibits a redshift, with a slight increase in the absolute value, as shown in Figure 5c. Finally, the energy-spreading analysis presented in Figure 5d suggests that while the mode volume remains constant, using DNH structures with a 200 nm hole diameter yields the maximum trapping energy when excited at 850 nm. Thus, a distinct spectral property of DNH structures emerges from these observations: varying the diameter of the nanoholes is a straightforward strategy to control the plasmonic resonance wavelength in the NIR and tailor the sensitivity for trapping at the desired wavelength.

**Hole with pillar**

We explored the inclusion of a pillar within each nanohole, as it has been shown that the plasmonic resonance can be tuned to trap 20 and 100 nm polystyrene nanospheres, with higher trap stiffness for the same input irradiance[66,67]. Figure 5j shows four SEM images of fabricated Double Nanoring (DNR) structures incorporating such pillars. Figure 5e shows the simulated transmission spectra for DNR structures with pillar diameters ranging from 0 nm (no pillar) to 150 nm in 10 nm increments. The hole diameter $\boldsymbol{\phi}_{\text{bot}}$ was fixed to 170 nm. In agreement with previous studies[67], the presence of large pillars broadens and red-shifts the plasmonic resonance while attenuating the transmission. This effect is explained by mode hybridization between the inner and outer ring, which shifts towards the NIR as the cavity narrows[68]. Interestingly, including pillars of diameters up to ~100 nm has no profound effect on $\Delta T/T_0$ as seen in Figure 5f. This aspect is also reflected in the small change of the zero-crossing point for those pillar diameters (Figure 5k). For larger diameters, the peak of $\Delta T/T_0$ is attenuated and redshifted, yielding a near-zero sensitivity over a long wavelength range, meaning that such DNR structures are ineffective for trapping in that spectral range.

Figure 5g shows the maximum electric-field enhancement spectra of DNR structures. As the pillar diameter increases, there is a slight improvement in the maximum enhanced field, along with a redshift. However, for pillar diameters above ~100 nm, the enhancement drops when compared to the no-pillar structure. Finally, as the pillar diameter increases, the rings become thinner, causing the DNR to narrow along its entire perimeter. Consequently, the localized field enhancement at the gap is no longer the sole hotspot within the cavity. Instead, the field confinement spreads uniformly around the DNR. This effect is reflected in a marked increase in the mode volume and a reduction in the mode energy for pillar diameters above 120 nm (50 nm-thin rings), as shown in Figure 5h.

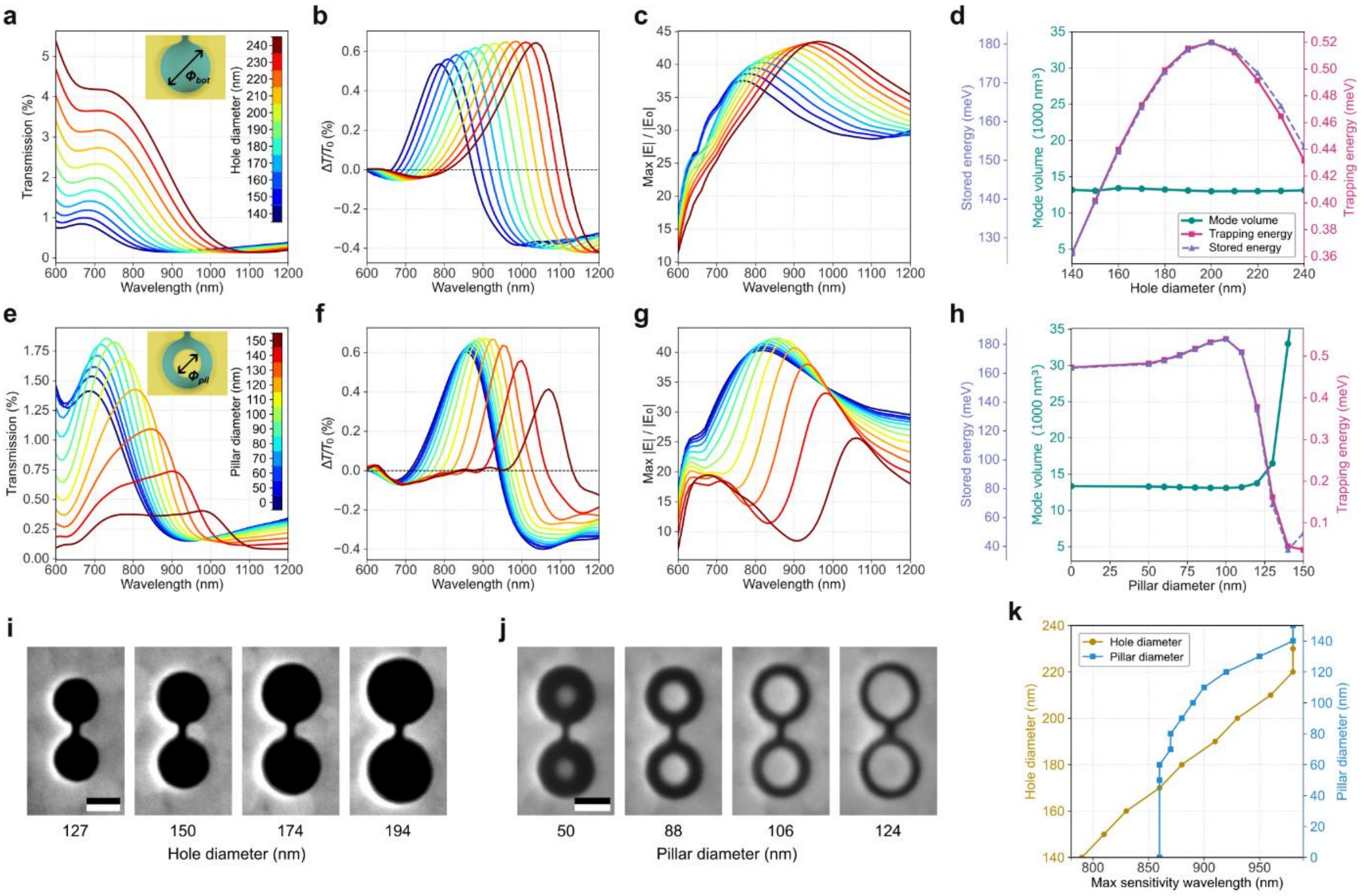

*Figure 5: Effect of DNH hole and pillar dimensions on the optical properties. (a and e) Transmission spectra for different hole and pillar diameters, respectively. (b and f) Corresponding transmission variation upon trapping $\Delta T/T_0$. (c and g) Corresponding maximum electric-field enhancement. (d and h) Corresponding stored energy, mode volume, and trapping energy. (i) SEM images (top view) of DNH structures fabricated with different hole diameters. (j) SEM images (top view) of DNH structures fabricated with pillars of different diameters. (k) Wavelength of maximum $\Delta T/T_0$ for the explored parameters.*

# Discussion

We have explored how the morphology of DNHs affects their optical properties through a comprehensive parametric study in which one geometric parameter is varied at a time. Among all the parameters considered, the thicknesses of the gold film and the titanium adhesion layer have the greatest impact on the spectral properties of the DNH structures. For instance, combining a 2 nm-thick Ti adhesion layer with a 200 nm-thick Au film could yield a substantial increase in both electric-field enhancement and sensitivity in the down-trap regime.

The DNH geometry also offers a few simple ways of tuning the resonance and sensitivity. Increasing the hole diameter makes it possible to redshift the sensitivity peaks toward a desired trapping wavelength. Additionally, modifying the profile of the wall taper can further increase the electric-field enhancement. The individual effect of each parameter is summarized in Supporting Information Table S3.

While independent parametric analyses clearly indicate those trends, exploring the combined parameter space may reveal designs that further optimize trapping performance. Focusing on the up-trapping of small proteins at 850 nm, we followed two complementary optimization strategies. In the first approach, we selected parameter values heuristically within the explored range, leveraging or compensating for spectral shifts to simultaneously improve sensitivity and electric-field enhancement. In the second, we employed a Differential Evolution algorithm[69] using the sensitivity as the figure of merit (see Methods for further details on the optimization process). This global optimization method treats the FDTD simulation as a black-box evaluator and samples the parameter space without requiring gradient computation of the sensitivity, which would be highly time- and resource-consuming. Its stochastic search strategy allows it to escape narrow local minima that plasmonic resonances are likely to introduce in the optimization landscape, which would otherwise stall gradient-descent or downhill simplex methods.

To ensure a feasible and experimentally realizable design, we set the gap bottom width to 18 nm in both cases. This dimension defines the maximum size of objects that can physically access the mode volume by diffusion: particles with a diameter exceeding the gap width are geometrically excluded (hard-sphere size limit). In practice, this constraint is further modified by electrostatic interactions between the object and the electrical double layer at the gold/fused silica interface, as well as by any surface coating, if present. For proteins such as Adenylate Kinase, Citrate Synthase, Hemoglobin, Calmodulin, and Ferritin, an effective gap width of a few times the protein size, approximately 10 – 25 nm based on their diameter, is sufficient to permit diffusion into the trapping volume, as previously reported[10,27,44] (see Supporting Information Section S6 for protein sizes). The remaining parameters were bound and conditioned as shown in Supporting Information Table S5. We did not include pillars inside the nanoholes in this optimization study.

*Table 2: Optimized DNH morphological parameters for trapping small proteins at 850 nm using a heuristic approach and a differential evolution algorithm. Original values correspond to the DNH structure of Figure 1.*

| | $\boldsymbol{t}_{\mathrm{Au}}$ (nm) | $\boldsymbol{t}_{\mathrm{Ti}}$ (nm) | $\boldsymbol{t}_{\mathrm{SiN}}$ (nm) | $\boldsymbol{g}_{\mathrm{w}}^{\mathrm{bot}}$ (nm) | $\boldsymbol{g}_{\mathrm{w}}^{\mathrm{top}}$ (nm) | Cusp convexity ($\mathrm{nm}^{-1}$) | $\boldsymbol{g}_{\mathrm{l}}$ (nm) | Wall taper | Etch depth (nm) | $\boldsymbol{\phi}_{\mathrm{bot}}$ (nm) | $\boldsymbol{\phi}_{\mathrm{top}}$ (nm) | $\boldsymbol{\phi}_{\mathbf{pil}}$ (nm) |
|---|---|---|---|---|---|---|---|---|---|---|---|---|
| **Original** | 100 | 5 | 30 | 18 | 100 | 0.0130 | 35 | 1 | 20 | 190 | 300 | 0 |
| **Heuristic** | 200 | 2 | 15 | 18 | 120 | 0.0415 | 20 | 2 | 40 | 240 | 280 | 0 |
| **Diff. evo.** | 200 | 2 | 15 | 18 | 108 | 0.0467 | 14 | 2.4 | 52 | 228 | 268 | 0 |

The optimized morphological parameters are presented in Table 2. Figure 6 shows the direct comparison between the optimized structures and the original DNH displayed in Figure 1. The transmission variation upon trapping is significantly higher at 850 nm in both approaches: 6.5 times more sensitive to up-trap events with the heuristic approach and 22 times more with the differential evolution approach. Likewise, the electric-field enhancement shows a 2.4-fold improvement for the heuristic and a 1.4-fold improvement for the differential evolution. Still, down-trapping is possible with both designs. For the heuristic design, at 920 nm, the electric-field enhancement is ~80, and the sensitivity is at its negative peak. Interestingly, the differential evolution design presents two options for down-trapping at ~740 nm and ~880 nm, with field enhancements of ~47 and ~68, respectively. Finally, the two optimized structures exhibit lower transmission than the original. This does not compromise the experiments, as detection in POT is typically performed with an avalanche photodiode, with neutral-density filters used to attenuate the transmitted light.

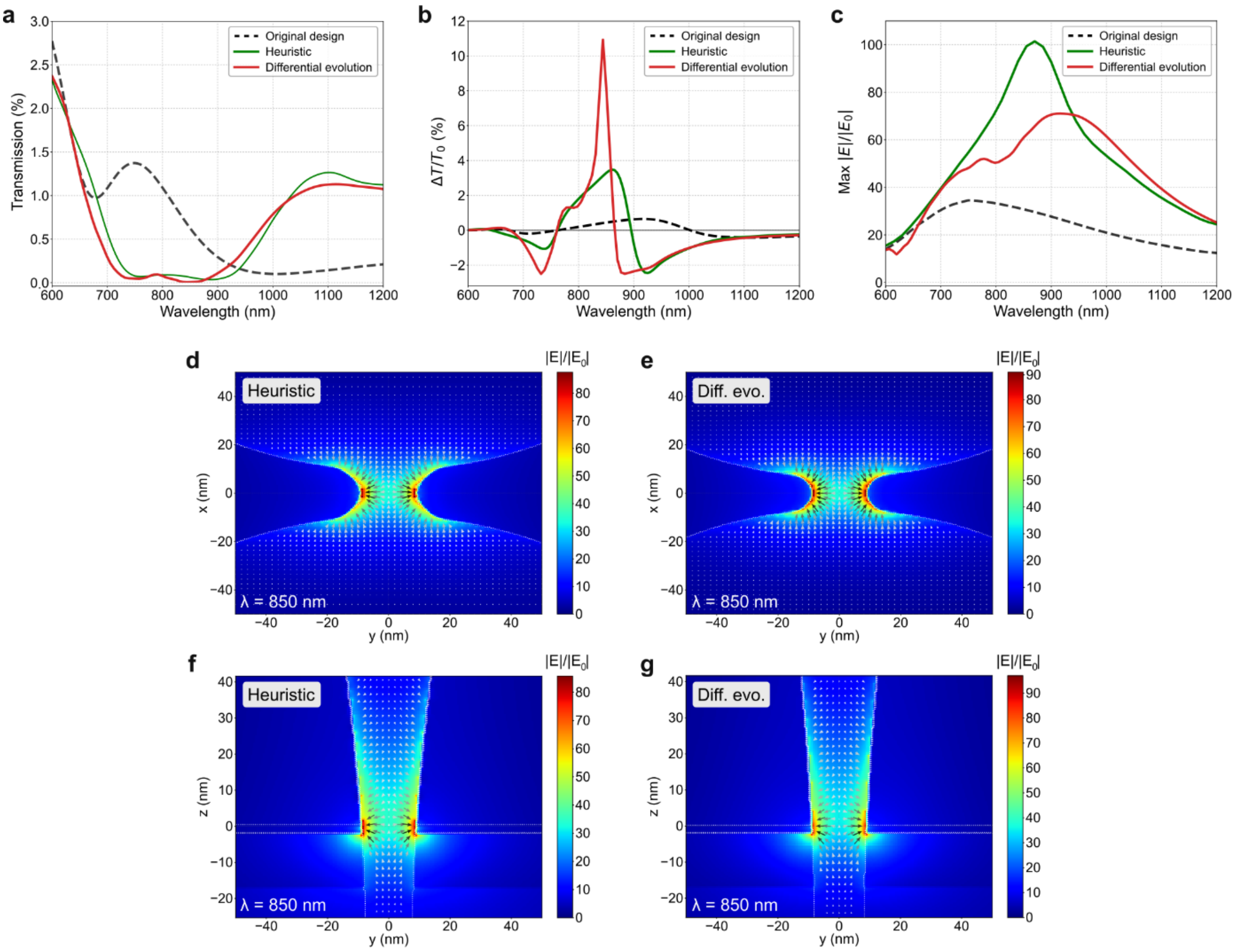

*Figure 6: Optical properties of the optimized DNH structures compared with the original DNH structure presented in Figure 1. (a, b, and c) Transmission, transmission variation upon trapping $\Delta T/T_0$, and maximum electric-field enhancement. (d and e) In-plane top view (xy) of the electric field in the region between the cusps, normalized by the incident electric field. The 2D field plots show the electric-field enhancement at 850 nm. The length of the arrows (in black and white for better contrast) indicates the magnitude and direction of the optical force acting on a particle as defined by Equation 1. (f and g) Same as panels d and f, but displaying the cross-sectional yz view of the electric-field enhancement at 850 nm.*

Figures 6d and 6e show the distribution of electric-field enhancements in the *xy* plane (top view, $z = 0$), and Figures 6f and 6g show the *yz* cross-section at the narrowest point of the gap ($x = 0$) for the two optimized configurations. The new designs exhibit highly confined electric fields with mode volumes of ~1000 nm$^3$ and ~700 nm$^3$ for the heuristic and differential evolution designs, respectively. These values represent reductions of roughly 12- and 17-fold, respectively, compared to the original DNH structure (~12000 nm$^3$). This strong spatial confinement results in large field gradients and, therefore, high optical forces. Such configurations are expected to yield higher trapping efficiencies, potentially enabling stable trapping at reduced laser powers.

In this work, we presented a systematic parametric study of how the morphology and layer composition of DNH structures govern the most relevant optical properties in plasmonic trapping, namely the transmission spectrum, its variation upon trapping, the local electric-field enhancement, and the mode volume of the plasmonic resonance. Rather than pursuing idealized geometries, we restricted each parameter to ranges that are realizable by FIB milling. We interpreted the results in light of practical considerations that commonly constrain Plasmonic Optical Tweezers, such as photothermal heating, long-term stability of the nanostructures, and the size of the object to be trapped relative to the available gap space.

We hope that these results will pave the way for the experimental community to validate and implement optimized DNH geometries under trapping conditions. Moreover, future work could focus on coupling electromagnetic simulations with a thermal model to directly quantify the photothermal effects associated with these designs. Finally, tailoring these fabrication-guided designs to specific applications, such as integration with solid-state nanopores for SANE sensing, would broaden the applicability of DNH-based plasmonic optical trapping for label-free single-protein characterization.

## Methods

### Electromagnetic simulations

Finite-Difference Time-Domain (FDTD) simulations were performed using Ansys Lumerical version 2025 R2. A plane wave with uniform amplitude was used as an incident electromagnetic field propagating in the $+z$ direction. The field was linearly polarized and perpendicular to the long axis of the DNH, i.e., parallel to the line that crosses both cusps. The substrate was placed in the $xy$ plane with the gold/titanium surface placed at $z$ = 0. The electric-field enhancement was computed by normalization to the incident electric field amplitude. To compute the transmission signal $T$, the transmitted electric field was measured with a power monitor located at $z$ = -800 nm and compared with the input power. We tested different spatial filters to represent the microscope objective's numerical aperture (NA), i.e., a 10× air objective with NA 0.25 and a working distance (WD) of 7.0 mm (CFI E-plan Achromatic, Nikon). We observed the Transmission variation upon trapping, $\Delta T/T_0$, is not affected by the spatial filtering (see Supporting Information Section S8 for comparison between different NAs). Hence, our simulations were performed without spatial filtering or far-field projection to reduce the computational load.

For the simulations with the 6 nm-diameter protein-like nanosphere, we used a refractive index of $\tilde{n}$ = 1.8 + j·0. The particle was positioned at the narrowest region of the gap, at a lateral distance of 7 nm from the gold sidewall, and vertically at the Ti/$SiN_x$ interface. This configuration emulates the space introduced by a commonly used antifouling self-assembled monolayer of PEG800. As a result, the surface-to-surface separation between the particle and the nearest gap sidewall was 1 nm.

We used the following refractive indices: water $\tilde{n}$ = 1.32 + j·0; fused silica substrate $\tilde{n}$ = 1.46 + j·0; silicon nitride $\tilde{n}$ = 2 + j·0; gold and titanium values were retrieved from Lumerical's database, selecting Johnson and Christy, and Palik tabulations, respectively.

The dimensions of the total simulated volume were 1 × 1 × 2 $\mu m^3$. We used a 1 nm mesh in the DNH area to simulate broadband Transmission, Transmission variation upon trapping, and Maximum electric-field enhancement. A finer mesh of 0.5 nm was used for single-wavelength simulations (850 nm) to compute the 2D electric-field plots, stored energy, trapping energy, and mode volume.

Parametric-sweep simulations were automated using the Lumerical Python API, with AI-assisted code generation. All simulation outputs were individually verified and subsequently post-processed for the analyses presented in this work.

The Differential Evolution algorithm was implemented using the SciPy Optimize library (version 1.16.1). The objective function was defined as the negative of the absolute value of the sensitivity, i.e., $-|\Delta T/T_0|$, ensuring that maximizing the sensitivity corresponds to minimizing the objective function. For each design candidate – a specific set of geometric parameters – the optimization procedure consisted of simulating the DNH transmission spectrum with and without the particle and subsequently evaluating the objective function at 850 nm. If the new candidate yielded an improved sensitivity, it replaced the parent and served as the basis for the next mutation. Otherwise, the parent was retained and further mutated. The optimization settings, including the mutation strategy, search bounds, and parameter-space discretization, are listed in Supporting Information Section S7. The optimization was terminated after 187 generations, and the best-performing design was selected from all evaluated candidates.

**Sample fabrication, characterization, and preparation**

First, 550 nm-thick fused silica wafers are cleaned with an RCA cleaning process. Then, a 30 nm-thick non-stoichiometric silicon nitride ($SiN_x$) layer is deposited onto the wafers by low-pressure chemical vapor deposition (LPCVD) at 800 $^{\circ}$C. Then, a 5 nm-thick titanium adhesion layer is deposited, followed by the deposition of a 100 nm-thick film of Au by electron-beam evaporation (Leybold Optics LAB 600H) at a substrate temperature of 190 °C at a rate of 4 Å/s and a wafer-target distance of 1010 mm. Later, the wafers are diced into 10 × 10 mm$^2$ dies (Disco DAD321). Finally, the nanostructures are milled into the multilayer using a focused ion beam (FIB, Thermo Scientific Scios 2 Dual Beam) with a gallium ion source operated at 30 keV and a current of 1.5 pA. DNHs were produced by milling, first, the two holes and, then, the rectangular region that defines the gap between them. To obtain the different tapered wall profiles, after milling the small rectangular area of the gap, a subsequent wider rectangular region was milled with a lower etching dose.

After fabrication, SEM images of pristine DNH structures were acquired in top-view and tilted modes using the SEM capabilities of the FIB system and a scanning electron microscope (Tescan Mira 3). SEM images were analyzed using ImageJ version 1.45p.

After characterization, the dies were mounted in a custom-made 3D-printed microfluidic chamber (see Supporting Information Section S9 for details of the chamber assembly and sample mounting). Finally, the chamber was filled with milli-Q water.

**DNH spectra acquisition**

The transmission spectra of as-fabricated DNHs were acquired in a custom-made inverted transmission confocal microscope (see simplified scheme in Supporting Information Section S10). A super-continuum broadband laser source (SuperK Fianium FIR-25, NKT Photonics) was used to illuminate the sample. The collimated beam was spectrally filtered with a short-pass filter (FESH1000, Thorlabs) and directed to a dichroic mirror via a 30R/70T non-polarizing cube beam splitter (BP020, Thorlabs). Then, the beam was sent to the excitation objective (CFI E Plan Achromat 100x, 1.25 NA, 0.23 WD, Nikon) using a dichroic mirror (T690spxxr, Chroma), resulting in a broadband illumination spanning the 690-1000 nm spectral range. Transmitted light was collected using an air objective (CFI E Plan 10x, 0.25 NA, 7 mm

WD, Nikon) and reflected towards a Czerny-Turner spectrometer with a linear CCD camera (Kymera 328i and Newton DU920P-BEX2-DD, Andor, Oxford Instruments) using a dichroic mirror (T510lpxrxt, Chroma). The optical setup included an LED and a CMOS camera (CS165CU1/M, Thorlabs) to inspect and locate the structures within the field of view. Fine positioning of the DNH at the laser focus was achieved using a 3-axis piezoelectric translation stage (NanoMax 311D/M, Thorlabs).

The laser was operated at 100% power to get the full supercontinuum spectrum. To avoid damaging the DNH structures, the power delivered to the sample was adjusted using a neutral-density filter with an OD of 3.0 (F06-925, Chroma), and the laser's pulse repetition rate was set to 0.15 MHz. Potential DNH damage was evaluated with dummy DNH structures under the SEM before and after laser irradiation.

To obtain the transmission spectrum of a DNH, $T_{\mathrm{DNH}}$, the transmitted light through the structure ($T_{\mathrm{sample}}$) was normalized by the spectrum of the light source ($T_{\mathrm{source}}$). The latter was acquired using a 5 µm-diameter aperture (large enough to avoid light interaction) milled in the same die as the DNHs, to account for transmission through the substrate and the chamber. In this way, $T_{\mathrm{DNH}}$ is proportional to $T_{\mathrm{sample}}/T_{\mathrm{source}}$. All spectra were acquired 3 times and later averaged. Each acquisition was bias-corrected and baseline-corrected to remove any stray-light contribution. Exposure times were adjusted based on measured intensities to avoid camera saturation. For the DNH, we used an integration time of 7 s per exposure. For the source, we used an integration time of 0.1 s. Therefore, the DNH spectrum shown in Figure 1i was computed considering the exposure times ratio, $T_{\mathrm{DNH}}(\%) = 100 \cdot \frac{T_{\mathrm{sample}}/70}{T_{\mathrm{source}}}$. Finally, the spectrum was arbitrarily scaled by a factor of 4.8 to compare with the simulated data. No correction was applied to compensate for the difference between the experimental and simulated numerical apertures.

## Code availability

Codes are available upon request.

## Acknowledgements

This research was supported by the Swiss National Science Foundation (SNSF), Grant No. 200020_197239 to M.M., the Adolphe Merkle Foundation, and the University of Fribourg. P.M. acknowledges the SNSF Swiss Postdoctoral Fellowships (Grant No. TMPFP3_217470). M.B. thanks A. Mukhopadhyay for fruitful discussions and O. Dyberg for his assistance during the measurements.

# Supporting Information

for

# Morphology-driven Optimization of Double Nanoholes for efficient and sensitive Plasmonic Optical Trapping

Pau Molet[1,2], Edona Karakaçi[1], Maria Sanz Paz[1,3], Mariano Barella[1,*], Michael Mayer[1,*]

[1]Adolphe Merkle Institute, University of Fribourg, Chemin des Verdiers 4, CH-1700 Fribourg, Switzerland.

[2]Present address: Department of Electrical Engineering, Stanford University, Stanford, CA 94305, USA.

[3]Present address: Sorbonne Université, CNRS, Institut des NanoSciences de Paris, INSP, F-75005 Paris, France.

*Corresponding authors:

mariano.barella@unifr.ch

michael.mayer@unifr.ch

## Table of Contents

## S1. Particle position in the SIBA mechanism

The presence of the nanoparticle or protein in the hotspot modifies the local dielectric function, and, consequently, the enhanced electric field differs from that of the empty gap. This local enhancement depends on the object's position within the gap. Figure S1 shows the electric-field enhancement for several nanoparticle positions (top row) together with the corresponding difference in enhancement, that is, the field of the DNH with the particle minus that of the empty gap (bottom row). The smaller the space between the object and the DNH surface, the higher the local enhancement.

Additionally, the magnitude of this change relative to the empty gap depends on the object's size. Figure S2 shows the electric-field enhancement for several nanoparticle diameters (top row) together with the corresponding difference in enhancement (bottom row). The distance between the nanoparticle surface and the DNH surface was fixed at 1 nm. The larger the object, the higher the local field enhancement: for example, a 12 nm-diameter nanoparticle within the gap increases the maximum local electric-field enhancement by ~20% (Figure S2c), whereas a 4 nm-diameter nanoparticle increases it by ~14% (Figure S2a).

Simulations were performed as shown in Figure 1 of the main text (see Methods for further details).

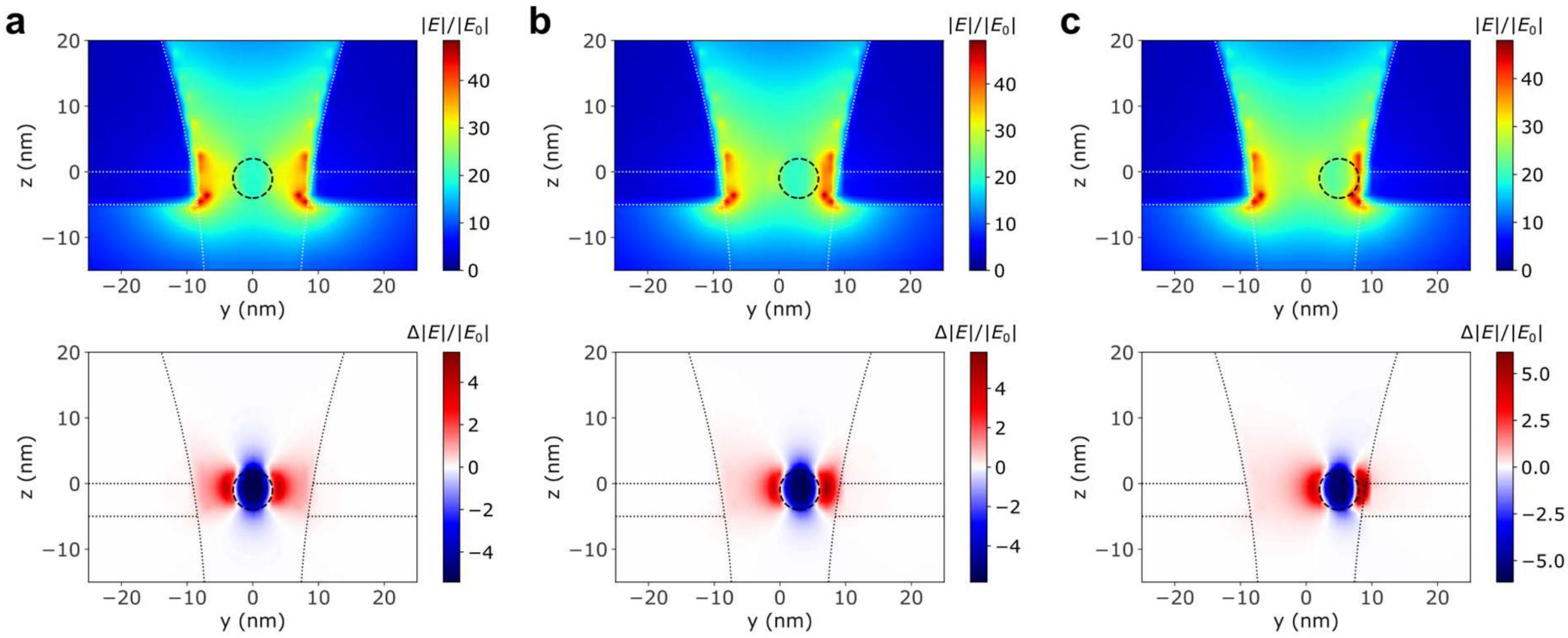

*Figure S1: Position dependence in the SIBA mechanism. The local field enhancement changes with the object's position. Top row: electric field enhancement in the gap. Bottom row: difference in electric field enhancement. The particle's center position is x = 0 nm and (a) y = 0 nm, (b) y = 3 nm, and (c) y = 5 nm.*

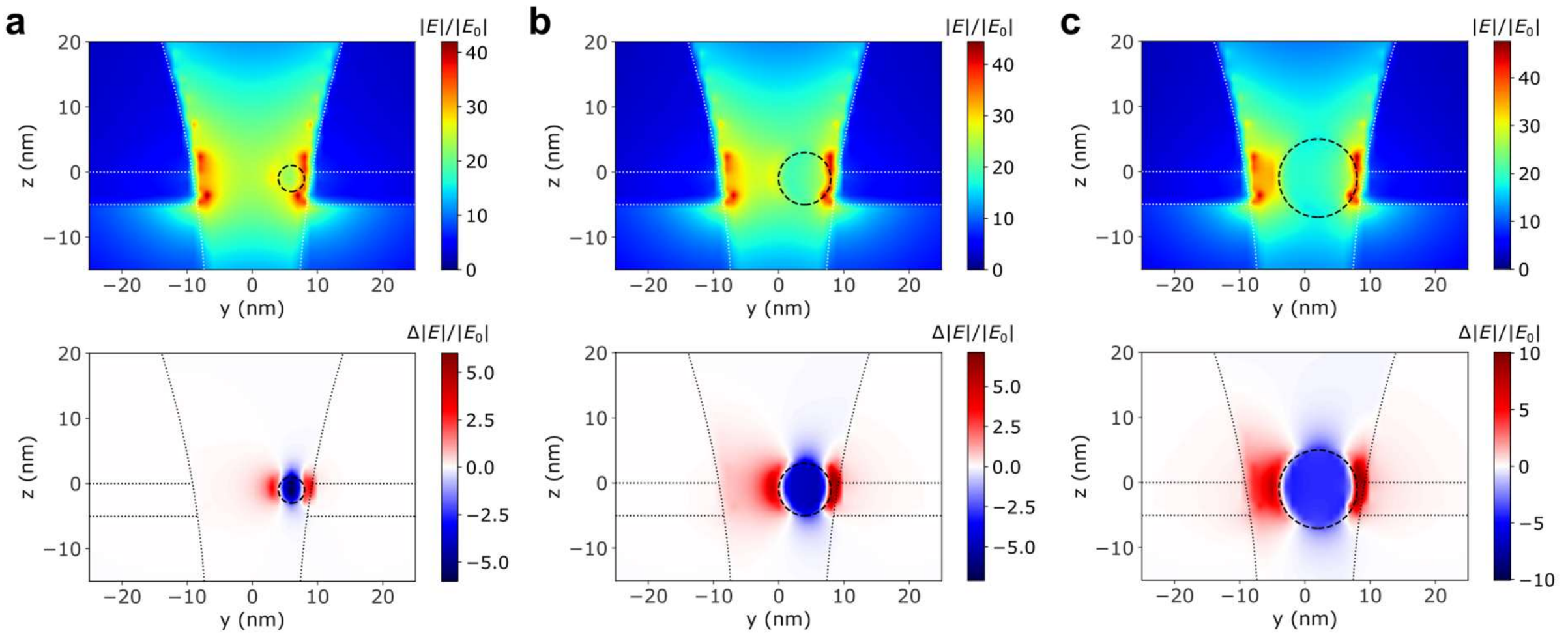

*Figure S2: Object size dependence in the SIBA mechanism. The local field enhancement changes with the object's size. Top row: electric field enhancement in the gap. Bottom row: difference in electric field enhancement. For a fixed object surface-DNH wall of 1 nm, the particle's radius is (a) 2 nm, (b) 4 nm, and (c) 6 nm.*

## S2. FIB-milled DNH structures morphology

To analyze the reproducibility and variability of our fabrication method, we studied four groups of 48 FIB-milled DNH structures. Each group had a fixed gap length. They were fabricated on a 100 nm-thick Au film, with 5 nm of Ti as an adhesion layer and 30 nm of silicon nitride. Refer to the Methods section of the main text for further details about the fabrication. All four groups were designed to have the same hole diameter and the same gap bottom width. Representative SEM images of each group are shown in Figure S3a.

To retrieve the morphological parameters, each DNH was imaged by SEM using identical acquisition settings (10 kV, 5 mm working distance, beam intensity 10, beam spot size of 5.8 nm, 200× magnification, In-beam secondary-electron detector). Representative SEM images of the different DNH are presented in Figure S3a. Beyond the manual measurements performed in ImageJ, we developed an automated analysis to extract the morphological parameters. Briefly, the code (available on GitHub) crops the image around the DNH, applies an intensity threshold set to the first minimum of the intensity histogram (see Figure S3b), smooths the resulting binary image, and extracts the continuous perimeter of the DNH. Circle and sine functions are then fitted to the holes and cusps, respectively, and the parameters are retrieved by measuring the DNH dimensions defined in Figure 1 of the main text. As each DNH contains two holes and two cusps, the average of the two is reported as the measured dimension.

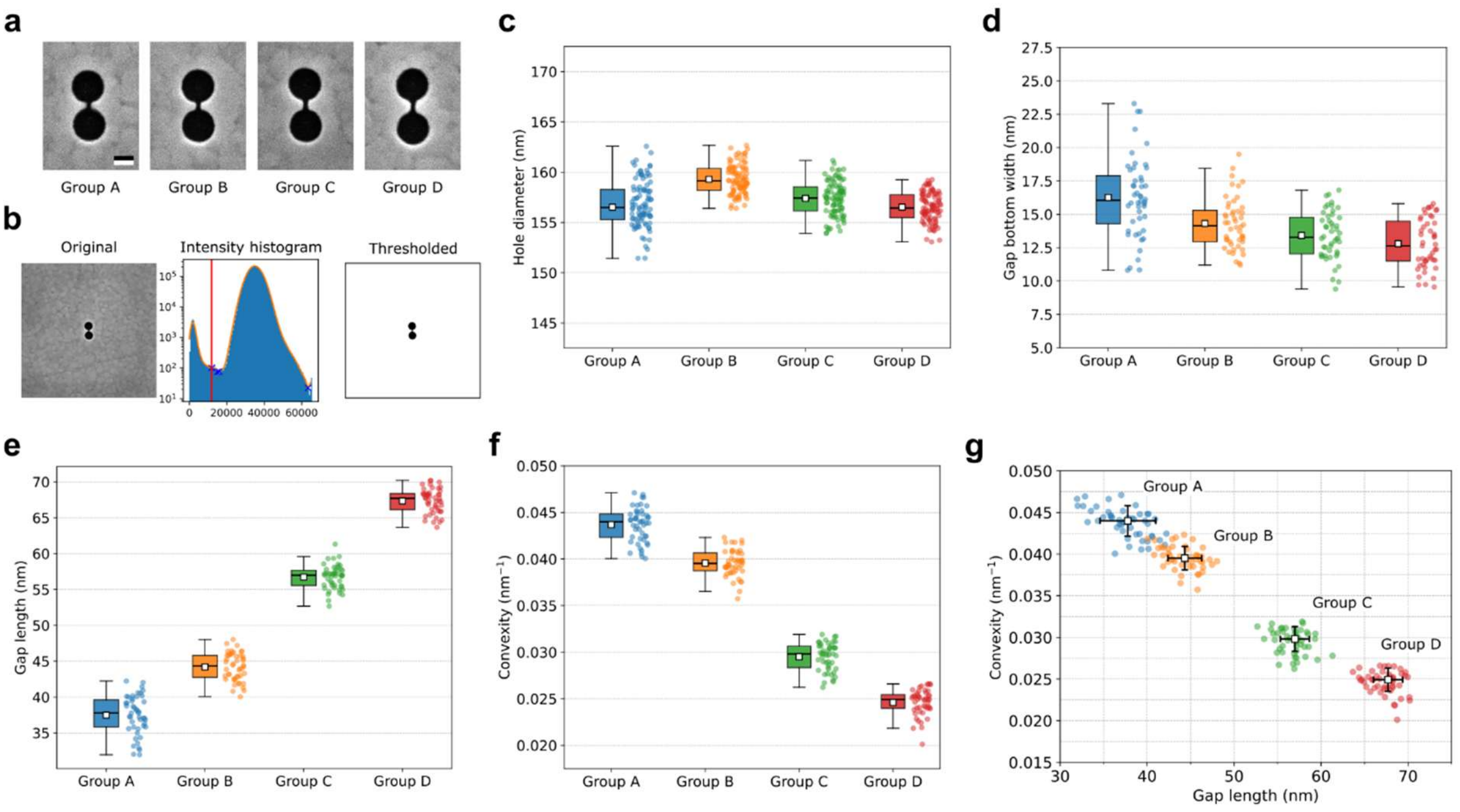

*Figure S3: Geometry of as-fabricated DNH structures with different gap lengths. Number of structures per group: 48. (a) Representative SEM images of one DNH per group. Images were acquired with the same imaging parameters. Scale bar: 100 nm. (b) Thresholding intensity method for image analysis. Parameters distribution and box plots of (c) Hole diameter, (d) Gap bottom width, (e) Gap length, and (f) Convexity. The boxes extend between the 25% and 75% percentiles. The black horizontal line is the dataset's median. The mean is represented by a white square symbol. (g) Geometrical link between convexity and gap length. The white square symbol represents the mean, and the black lines are one standard deviation.*

Table S1 reports the median and the standard deviation of the retrieved parameters for the four groups. In all cases, the hole diameter shows the lowest variability, at 0.6% (defined as the standard deviation divided by the median). Then, the convexity follows at 5%, the gap length at 8%, and the gap width with variations of up to 15%. These trends are illustrated in Figures S3c-f, which show low fabrication variability across most of the parameters. The exception is the gap width, whose dispersion increases as the gap narrows, indicating that FIB milling becomes less reproducible when fabricating the smallest features. Figures S3e and S3f further reveal a relationship between convexity and gap length, and Figure S3g shows that this relationship is approximately linear.

*Table S1: Mean and standard deviation DNH structures' morphological parameters. Group size = 48.*

| Group | Hole diameter (nm) | Gap bottom width (nm) | Gap length (nm) | Convexity ($nm^{-1}$) |
|---|---|---|---|---|
| A | 156 ± 1 | 16 ± 3 | 38 ± 3 | 0.044 ± 0.002 |
| B | 159 ± 1 | 14 ± 2 | 44 ± 2 | 0.040 ± 0.002 |
| C | 157 ± 1 | 13 ± 2 | 57 ± 2 | 0.030 ± 0.002 |
| D | 156 ± 1 | 13 ± 2 | 68 ± 2 | 0.025 ± 0.002 |

Another aspect of FIB-milled DNHs is the rounded edges on the gold top surface. This feature is visible only when SEM images are acquired with a tilted sample. Figure S4 shows this fabrication effect in different DNHs.

a

b

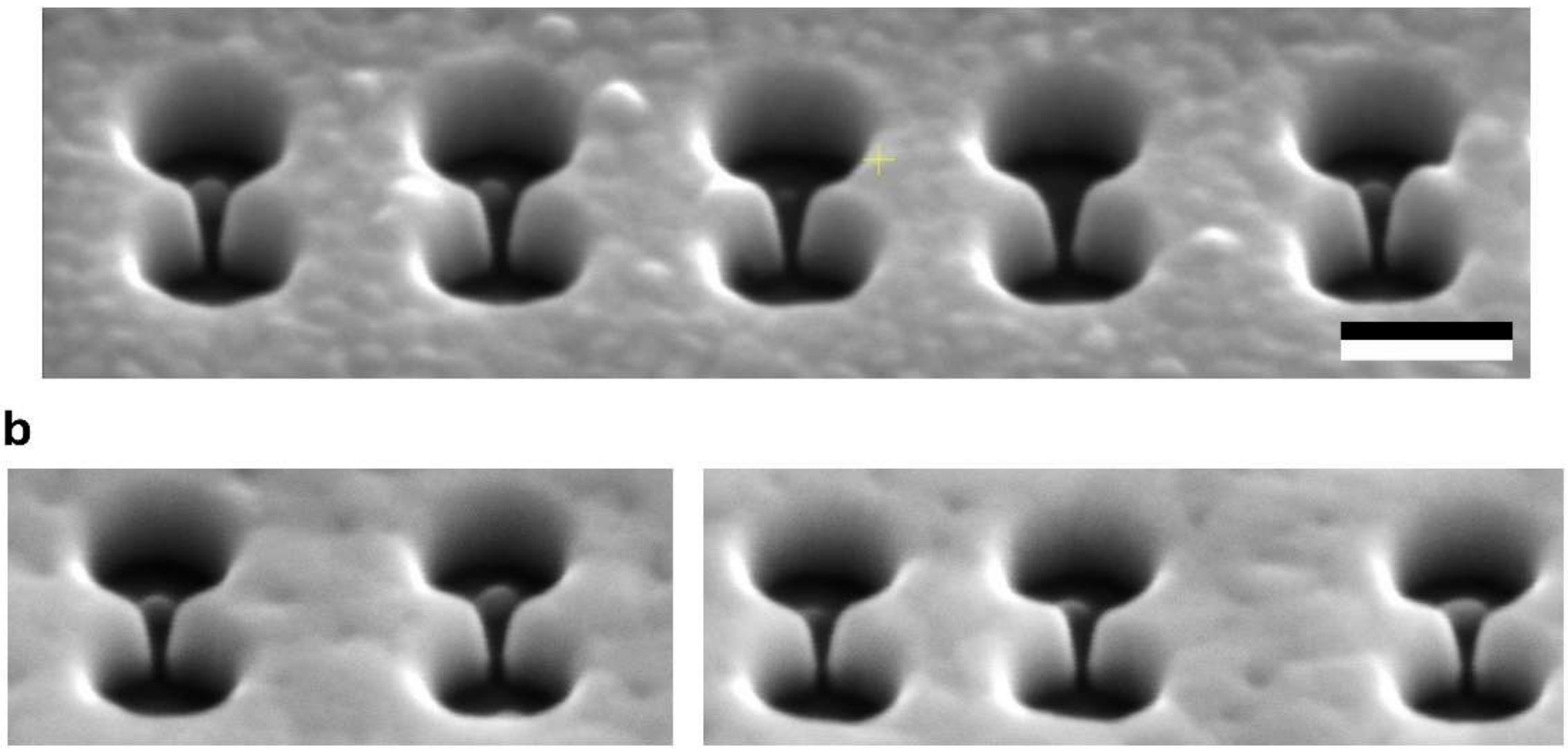

*Figure S4: Round edges of the gold top surface of as-fabricated DNH structures with different gap lengths on a $SiN_x$/Ti/Au stack. Gold deposited at (a) room temperature, and (b) 190° C. Both panels share the same scale bar of 200 nm.*

## S3. Gold thin-film grain size of different deposition methods

Figure S5 shows SEM images of various gold thin films deposited on different material stacks under different conditions. The fabrication of the reference material stack ("REF") is described in the Methods section of the main text. Briefly, the stack is deposited onto a 550 µm-thick fused silica wafer and consists of a $SiN_x$ layer, a Ti adhesion layer, and an Au film.

We evaluated the effect of alternative fabrication strategies on the gold grain size. They include: a self-assembled monolayer based on (3-Mercaptopropyl)trimethoxysilane (MPTMS, 4420-74-0, Sigma-Aldrich) as a replacement for the Ti adhesion layer, the absence of silicon nitride, a thinner titanium adhesion layer, and different substrate temperatures during gold deposition (room temperature or high temperature).

For the functionalization of the fused silica surface with MPTMS, we followed the process described by Riccardi and coworkers[1]. In short, the fused silica wafer underwent RCA cleaning, oxygen plasma activation, and overnight exposure to the vapor phase of 1 ml of MPTMS in a desiccator. Then, the wafer was baked at 80 °C on a hot plate for 1 h and, finally, rinsed with isopropanol before gold deposition.

For the gold deposition at room temperature, we used the electron-beam evaporator EVA 760 (Alliance Concept) with a wafer-target distance of 350 mm and a deposition rate of 1 Å/s. Refer to the Methods section for further details.

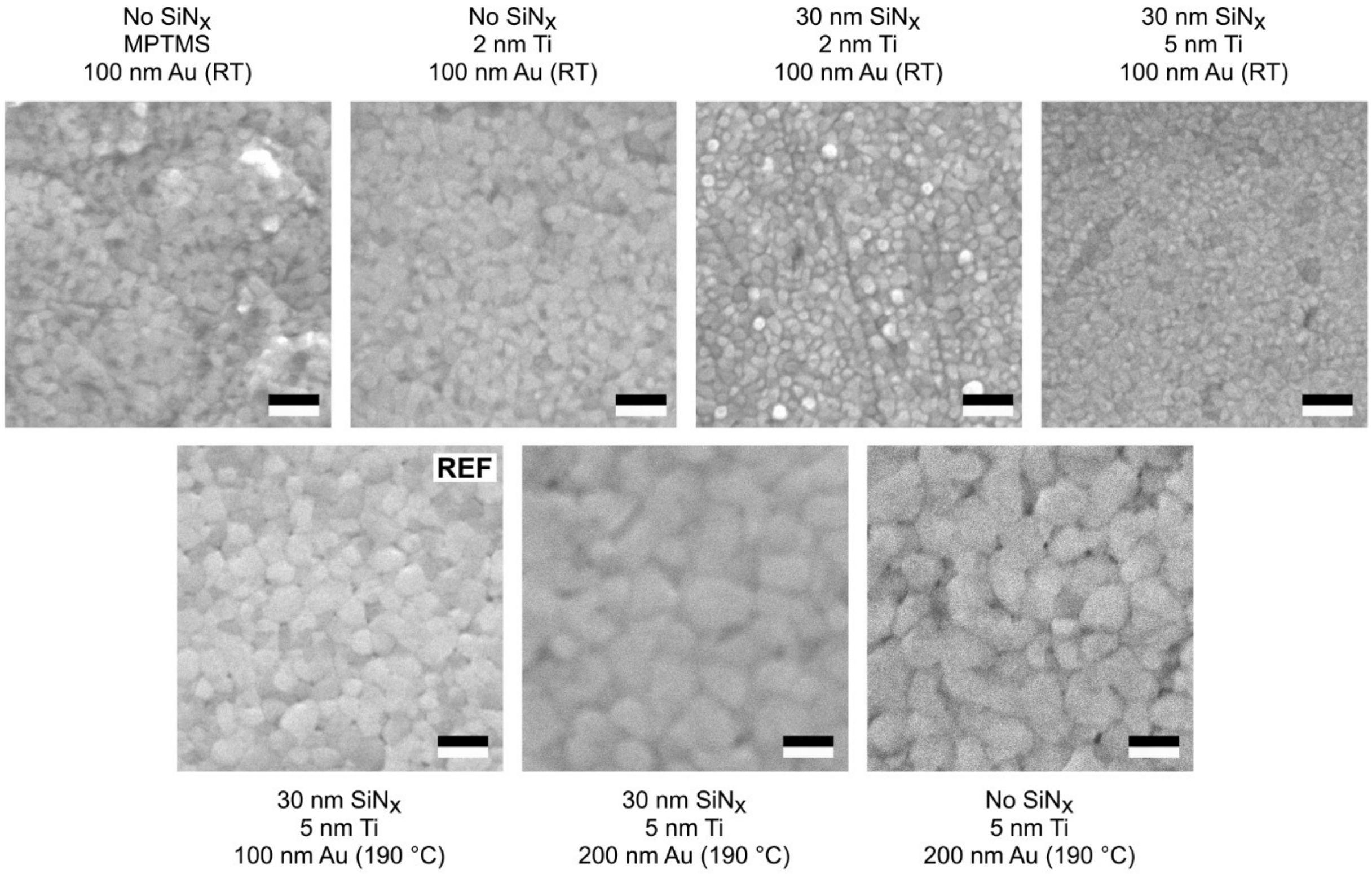

*Figure S5: SEM images of different gold thin films deposited on different substrates and under different conditions. Our reference material stack is indicated as "REF". Scale bar: 200 nm.*

The grain size distributions for the samples shown in Figure S5 are presented in Figure S6. A comparative analysis across these substrates leads us to the following conclusions: the temperature during Au deposition plays a crucial role in determining the grain size; the thickness of the titanium adhesion layer has a small impact on the grain size; the presence of silicon nitride layer causes smaller grain sizes; the use of MPTMS as an adhesion layer yields a gold grain size similar to that obtained with a 2 nm titanium adhesion layer.

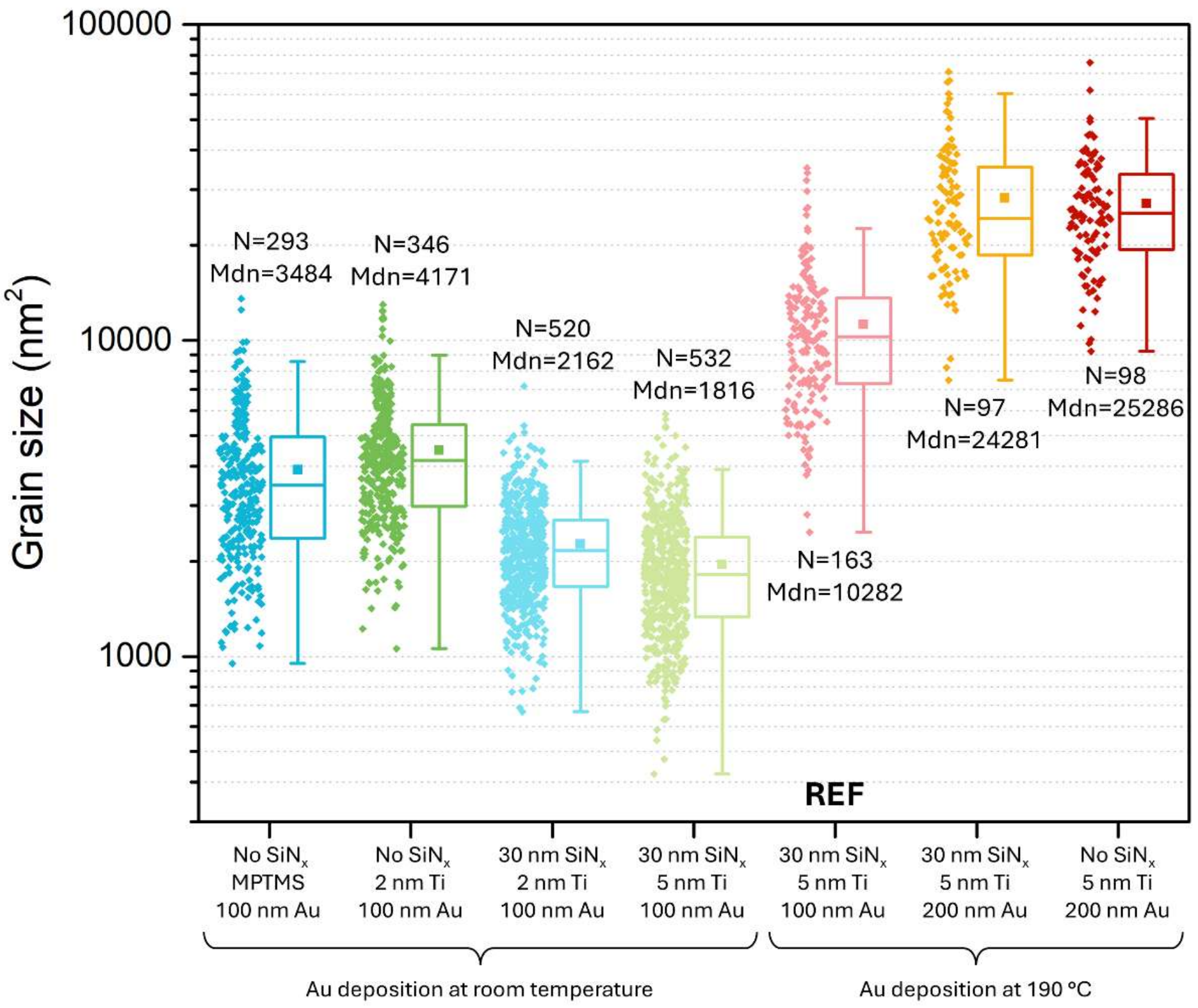

*Figure S6: Box plots of the grain size distribution in gold thin films deposited under different temperatures and conditions. Each box spans the 25$^{th}$ and 75$^{th}$ percentile range, with the horizontal line indicating the median. The mean grain size is indicated by a solid square. The text adjacent to each box plot indicates the sample size (N) and the median (Mdn). Our reference material stack is indicated as "REF".*

## S4. Linked geometry of tapered-gap DNH simulations

Creating tapered walls on a DNH using FIB milling affects both the top width of the gap and the top diameter of the hole in the same way. The increase in the gap at the top and of the hole at the top are linked as described in Table S2. The hole bottom diameter is set to 170 nm as described in the Simulation setup subsection of the main text. The tapering profile was set to $N = 1$.

*Table S2: Gap top width and the corresponding hole top diameter in tapered-gap DNHs.*

| | | | | | | | | | | |
|---|---|---|---|---|---|---|---|---|---|---|
| **Gap top width (nm)** | 18 | 30 | 50 | 70 | 90 | 110 | 130 | 150 | 170 | 190 |
| **Hole top diameter (nm)** | 170 | 180 | 190 | 200 | 210 | 220 | 230 | 240 | 250 | 270 |

## S5. Effect of each DNH morphological parameter on the studied optical properties

The individual effect of each parameter is summarized in Table S3. A column that reports on the difficulty of reproducing a target or optimized value during FIB milling fabrication was added. For example, the narrowest gaps are the least reproducible, showing the highest variability of all measured features (see Section S2).

*Table S3: Brief description of the parameter's role on the DNH structure optical properties. Fabrication complexity ranges from Low to Moderate to High. Thin-film layer deposition presents no fabrication difficulty.*

| Parameter | Effect (while increasing) | Fabrication complexity |
|---|---|---|
| $\boldsymbol{t}_{\mathrm{Au}}$ (nm) | - Blueshifts the sensitivity<br>- Increases the field enhancement<br>- Better heat dissipation<br>- Increases the mode volume | - |
| $\boldsymbol{t}_{\mathrm{Ti}}$ (nm) | - Reduces the transmission, sensitivity, and enhancement<br>- Dampens the plasmonic effect | - |
| $\boldsymbol{t}_{\mathrm{SiN}}$ (nm) | - Slightly increases the transmission<br>- Reduces the sensitivity and the field enhancement<br>- No effect on mode volume | - |
| $\boldsymbol{g}_{\mathrm{w}}^{\mathrm{bot}}$ (nm) | - Blueshifts the sensitivity<br>- Cancels sensitivity and reduces the field enhancement<br>- Offers more space with a reduced mode volume | Moderate |
| Cusp convexity ($\mathrm{nm}^{-1}$) | - No significant impact on the optical properties<br>- Slight increase in the field enhancement<br>- Almost constant mode volume | Moderate/High |
| $\boldsymbol{g}_{\mathrm{l}}$ (nm) | - Slight increase in transmission<br>- Reduces the sensitivity and the field enhancement<br>- Coupled with the cusps' convexity | Low |
| Wall taper | - Reduces the transmission, sensitivity, and enhancement<br>- Slight increase in the mode volume | High |
| $\boldsymbol{g}_{\mathrm{w}}^{\mathrm{top}}$ (nm) | - Increases the transmission<br>- Blueshifts the sensitivity and the field enhancement<br>- Offers a minimum for the mode volume | High |
| Etch depth (nm) | - No impact on the optical properties<br>- Slight increase in the field enhancement<br>- Offers more space with an almost constant mode volume | Moderate |
| $\boldsymbol{\phi}_{\mathrm{bot}}$ (nm) | - Increases transmission<br>- Blueshifts the sensitivity and the field enhancement<br>- No effect on the mode volume | Low |
| $\boldsymbol{\phi}_{\mathrm{top}}$ (nm) | Not explored | High |
| $\boldsymbol{\phi}_{\mathrm{pil}}$ (nm) | - Reduces and blueshifts the transmission, the sensitivity, and the field enhancement for large diameters<br>- Increases the mode volume and delocalizes the hotspot | Low |

## S6. Protein size survey

The size of a protein can be estimated by[2]:

$$r = 0.066M^{1/3} \tag{S1}$$

where $M$ is the mass of the protein in Daltons and $r$ the radius in nanometers. In general, this equation underestimates the true size of the protein, as it assumes a perfectly spherical shape with a uniform density of 1.37 g/cm$^3$. Therefore, the values displayed in Table S4 should be taken as the minimum protein diameter.

*Table S4: Mass and diameter of proteins previously studied by plasmonic optical trapping estimated by Eq. S1.*

| Protein | Mass (kDa) | Diameter (nm) | Ref. for mass / Ref. POT study |
|---|---|---|---|
| Calmodulin | 16.8 | 3.4 | UniProt (P62157) / [3] |
| Adenosine Kinase | 23.7 | 3.8 | AK1 human, SRP6121, Sigma-Aldrich / [4] |
| Hemoglobin | 64.5 | 5.3 | [4] / [4] |
| Citrate Synthase | 85 | 5.8 | C3260, Sigma-Aldrich / [4] |
| Apo-Ferritin | 443 | 10.1 | Equine spleen, A3660, Sigma-Aldrich / [5,6] |

## S7. Differential Evolution optimization settings

The optimization method was implemented using the scipy.optimize.differential_evolution function with the following settings: strategy set to 'best1bin', number of parameters to be optimized was 8 (population size), mutation constant was randomly sampled from a continuous uniform distribution over the half-open interval [0.5,1.0), and recombination constant set to 0.7. The search bounds and the range discretization (steps) of the parameter space are shown in Table S5. We refer the reader to the SciPy documentation for further details on the algorithm[7].

*Table S5: Optimization search bounds and steps used in the Differential Evolution algorithm.*

| Parameter | Search bounds (min to max) | Step |
|---|---|---|
| $\boldsymbol{t}_{\mathrm{Au}}$ (nm) | 100 to 300 | 10 |
| $\boldsymbol{t}_{\mathrm{Ti}}$ (nm) | Fixed (2) | - |
| $\boldsymbol{t}_{\mathrm{SiN}}$ (nm) | 7.5 to 22.5 | 0.75 |
| $\boldsymbol{g}_{\mathrm{w}}^{\mathrm{bot}}$ (nm) | Fixed (18) | - |
| Cusp convexity ($nm^{-1}$) | 0.0208 to 0.0623 | 0.0021 |
| $\boldsymbol{g}_{\mathrm{l}}$ (nm) | 10 to 30 | 1 |
| Wall taper | 1.0 to 3.0 | 0.1 |
| $\boldsymbol{g}_{\mathrm{w}}^{\mathrm{top}}$ (nm) | 60 to 180 | 6 |
| Etch depth (nm) | 20 to 60 | 2 |
| $\boldsymbol{\phi}_{\mathrm{bot}}$ (nm) | 120 to 360 | 12 |
| $\boldsymbol{\phi}_{\mathrm{top}}$ (nm) | Fixed ($\boldsymbol{\phi}_{\mathrm{bot}}$ + 40) | - |
| Inside pillar | None | - |

## S8. Spatial filtering in the simulated transmission detection

We introduced spatial filters in the power monitor that collects the transmitted light to mimic different experimental conditions, that is, different microscope objectives. Figure S7a shows the simulated transmission signal for the same DNH, but different spatial filters were introduced into the power monitor to imitate objectives with numerical apertures (NA) of 0.10, 0.25, 0.60, and 1.00. As expected, the transmission signal decreases with NA due to reduced light collection. The spectral shape of the transmission is slightly perturbed by the NA, as shown with the Normalized transmission in Figure S7b. Lastly, the Transmission variation upon trapping, $\Delta T/T_0$, presents no significant differences with varying NA (Figure S7c). Following these observations, no spatial filter was applied in our simulations to reduce the computational load of filtering and far-field projection.

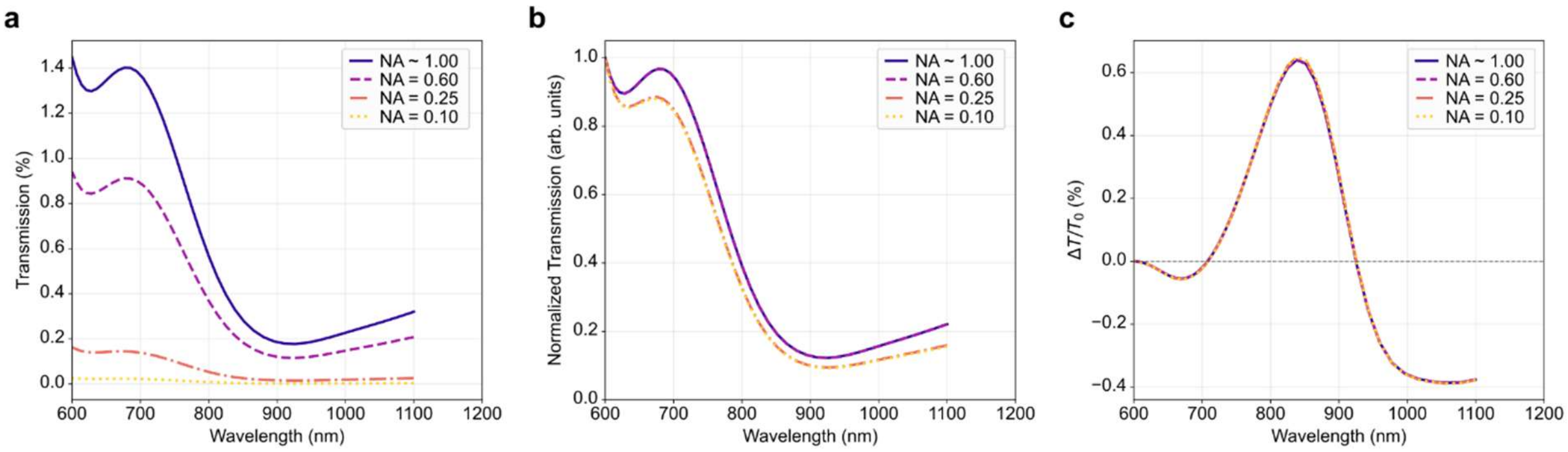

*Figure S7: Effect of the objective's NA. (a) Transmission, (b) Normalized transmission, and (c) Transmission variation upon trapping for NA 0.10, 0.25, 0.60, and 1.00.*

## S9. Microfluidic chamber assembly

Figure S8 shows the sample mounting and assembly of the microfluidic chamber. First, the sample was mounted in a custom-made 3D-printed microfluidic chamber. The die was glued to the chamber using a two-component silicone glue (Elite Double 16, Zhermack). A #1.5 borosilicate glass coverslip (Menzel Epredia) was subsequently rinsed with ethanol and isopropanol, and dried with compressed nitrogen. The coverslip was placed on top of the chamber and sealed with the glue. The distance between the gold surface containing the DNHs and the coverslip was ~80 µm. Finally, the tubes and the fittings were installed, and the chamber was filled with milli-Q water.

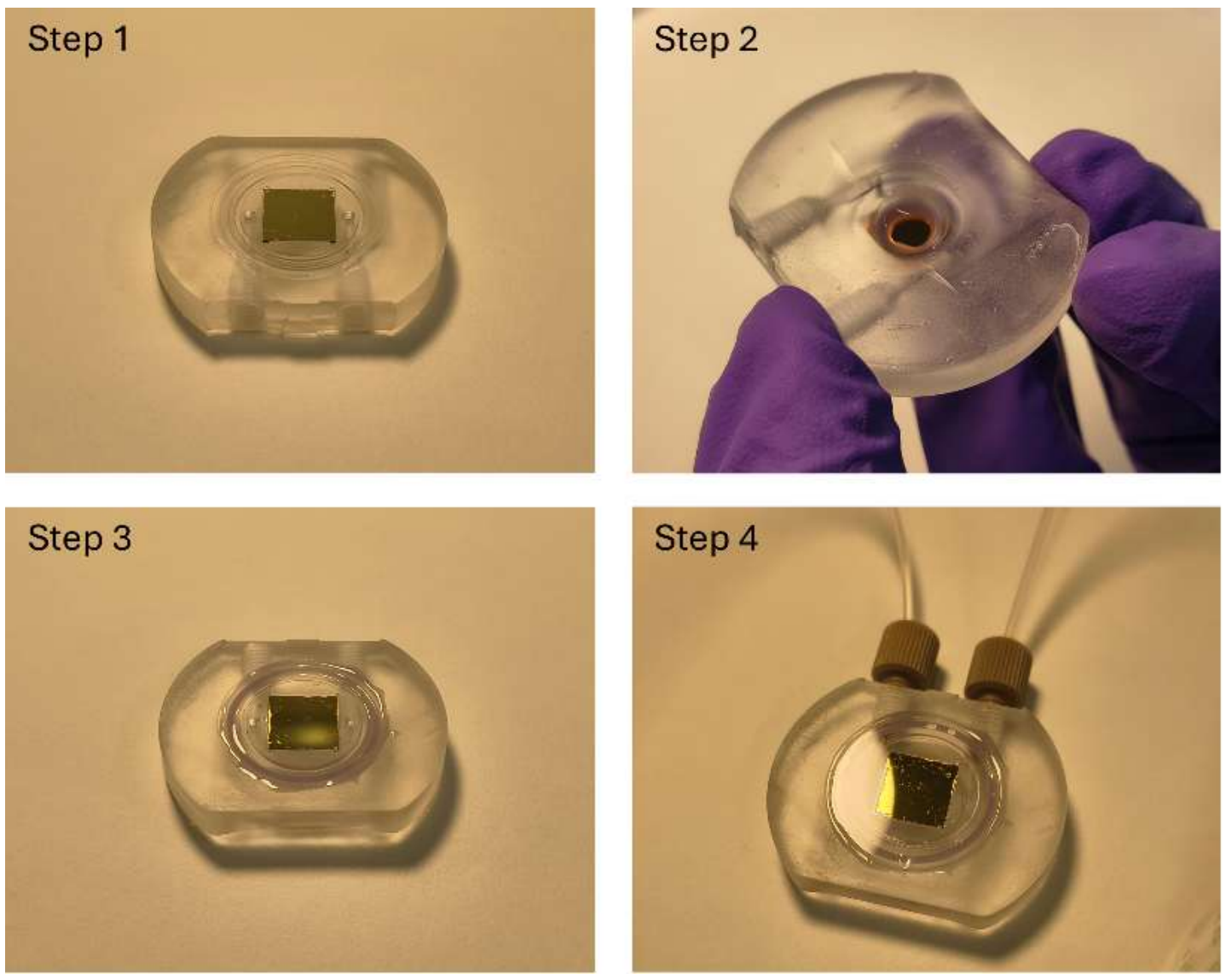

*Figure S8: Sample mounting and chamber assembly using a custom-made 3D-printed microfluidic chamber.*

## S10. Optical setup

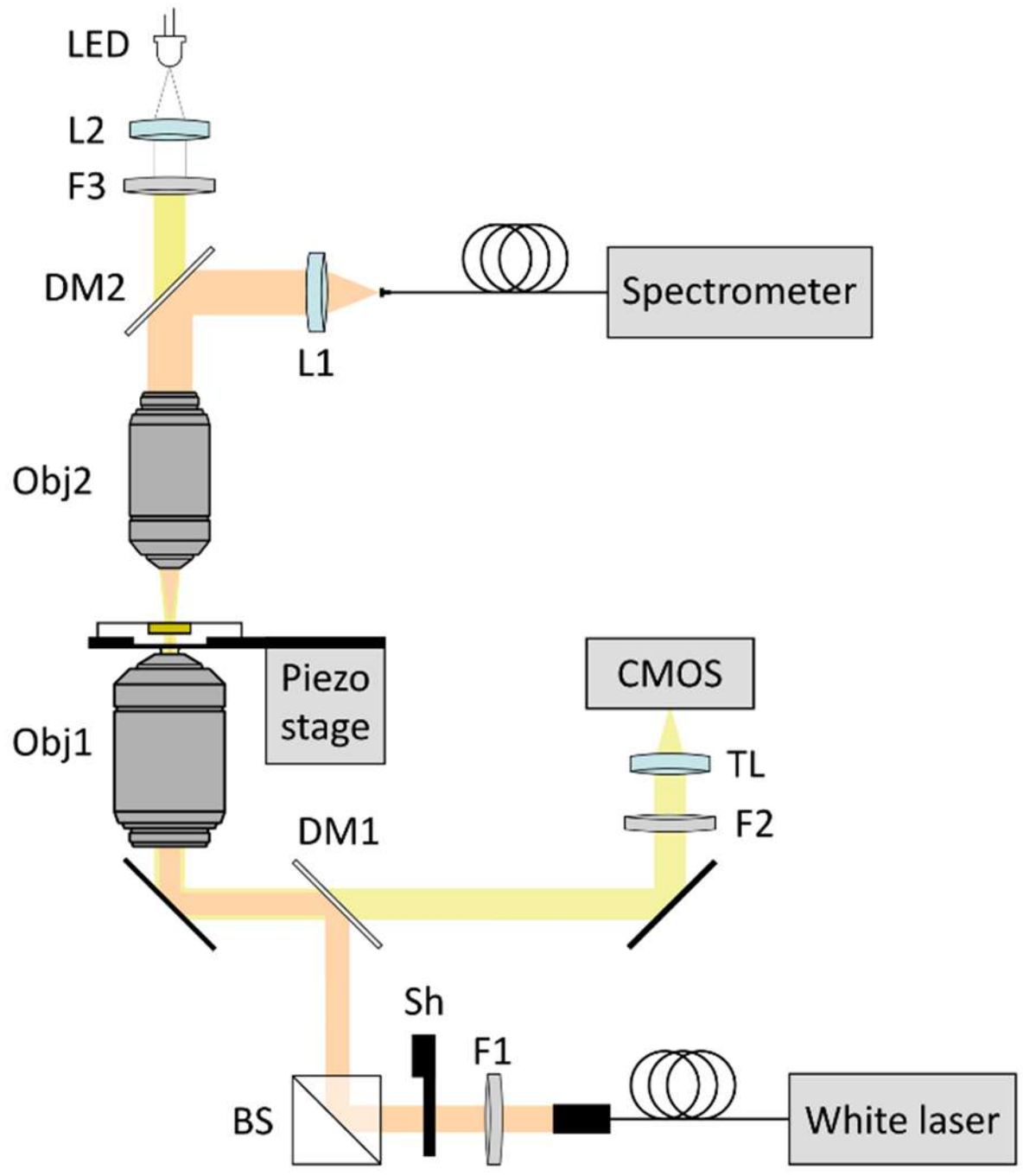

*Figure S9: Simplified scheme of the custom-made inverted transmission confocal microscope. F1: short-pass filter (FESH1000, Thorlabs) and neutral-density filter (UVND OD 3.0 V2, F06-925, Chroma). Sh: shutter (SH05R, Thorlabs). BS: non-polarizing beam splitter 30R/70T (BP020, Thorlabs). DM1: dichroic mirror (T690spxxr, Chroma). F2: NENIR40A, Thorlabs. TL: 200 mm-focal-length tube lens (AC254-200-A-ML, Thorlabs). Obj1: excitation objective. Obj2: Collection objective. DM2: dichroic mirror (T510lpxrxt, Chroma). F3: band-pass filter 565-595 nm (580/30, Omega). L2: ACL25416U-A, Thorlabs. LED: LEDW7E with LEDMT1E, Thorlabs. L1: 50 mm-focal-lens focusing lens (AC254-50-B-ML, Thorlabs).*